# *Euclid* preparation. Estimating galaxy physical properties using `CatBoost` chained regressors with attention

Euclid Collaboration: A. Humphrey⋆[1,2], P. A. C. Cunha[3,1], L. Bisigello[4,5], C. Tortora[6], M. Bolzonella[7], L. Pozzetti[7], M. Baes[8], B. R. Granett[9], A. Amara[10], S. Andreon[9], N. Auricchio[7], C. Baccigalupi[11,12,13,14], M. Baldi[15,7,16], S. Bardelli[7], C. Bodendorf[17], D. Bonino[18], E. Branchini[19,20,9], M. Brescia[21,6,22], J. Brinchmann[1], S. Camera[23,24,18], V. Capobianco[18], C. Carbone[25], J. Carretero[26,27], S. Casas[28], M. Castellano[29], G. Castignani[30,7], S. Cavuoti[6,22], A. Cimatti[31], C. Colodro-Conde[32], G. Congedo[33], C. J. Conselice[34], L. Conversi[35,36], Y. Copin[37], F. Courbin[38], H. M. Courtois[39], A. Da Silva[40,41], H. Degaudenzi[42], G. De Lucia[12], J. Dinis[40,41], F. Dubath[42], X. Dupac[36], S. Dusini[43], M. Farina[44], S. Farrens[45], S. Ferriol[37], M. Frailis[12], E. Franceschi[7], S. Galeotta[12], K. George[46], B. Gillis[33], C. Giocoli[7,47], A. Grazian[48], F. Grupp[17,46], L. Guzzo[49,9,50], S. V. H. Haugan[51], W. Holmes[52], I. Hook[53], F. Hormuth[54], A. Hornstrup[55,56], K. Jahnke[57], B. Joachimi[58], E. Keihänen[59], S. Kermiche[60], A. Kiessling[52], M. Kilbinger[45], B. Kubik[37], M. Kümmel[46], M. Kunz[61], H. Kurki-Suonio[62,63], S. Ligori[18], P. B. Lilje[51], V. Lindholm[62,63], I. Lloro[64], G. Mainetti[65], D. Maino[49,25,50], E. Maiorano[7], O. Mansutti[12], O. Marggraf[66], K. Markovic[52], M. Martinelli[29,67], N. Martinet[68], F. Marulli[30,7,16], R. Massey[69], H. J. McCracken[70], E. Medinaceli[7], S. Mei[71], M. Melchior[72], Y. Mellier[73,70], M. Meneghetti[7,16], E. Merlin[29], G. Meylan[38], M. Moresco[30,7], L. Moscardini[30,7,16], E. Munari[12], R. Nakajima[66], S.-M. Niemi[74], J. W. Nightingale[75,76], C. Padilla[26], S. Paltani[42], F. Pasian[12], K. Pedersen[77], V. Pettorino[74], S. Pires[45], G. Polenta[78], M. Poncet[79], L. A. Popa[80], F. Raison[17], R. Rebolo[32,81], A. Renzi[5,43], J. Rhodes[52], G. Riccio[6], E. Romelli[12], M. Roncarelli[7], E. Rossetti[15], R. Saglia[46,17], Z. Sakr[82,83,84], A. G. Sánchez[17], D. Sapone[85], R. Scaramella[29,67], P. Schneider[66], T. Schrabback[86], M. Scodeggio[25], A. Secroun[60], E. Sefusatti[12,14,13], G. Seidel[57], S. Serrano[87,88,89], C. Sirignano[5,43], L. Stanco[43], J. Steinwagner[17], P. Tallada-Crespí[90,27], A. N. Taylor[33], I. Tereno[40,91], R. Toledo-Moreo[92], F. Torradeflot[27,90], I. Tutusaus[83], L. Valenziano[7,93], T. Vassallo[46,12], A. Veropalumbo[9,20], Y. Wang[94], J. Weller[46,17], G. Zamorani[7], J. Zoubian[60], E. Zucca[7], A. Biviano[12,14], A. Boucaud[71], E. Bozzo[42], C. Burigana[4,93], M. Calabrese[95,25], R. Farinelli[7], N. Mauri[31,16], V. Scottez[73,96], M. Tenti[16], M. Viel[14,12,11,13,97], M. Wiesmann[51], Y. Akrami[98,99], V. Allevato[6], S. Anselmi[43,5,100], M. Ballardini[101,7,102], A. Blanchard[83], S. Borgani[103,14,12,13], S. Bruton[104], R. Cabanac[83], A. Calabro[29], G. Cañas-Herrera[74,105], A. Cappi[7,106], C. S. Carvalho[91], T. Castro[12,13,14,97], K. C. Chambers[107], S. Contarini[17,30], A. R. Cooray[108], J. Coupon[42], O. Cucciati[7], G. Desprez[109], A. Díaz-Sánchez[110], S. Di Domizio[19,20], J. A. Escartin Vigo[17], S. Escoffier[60], A. G. Ferrari[31,16], P. G. Ferreira[111], I. Ferrero[51], F. Fornari[93], L. Gabarra[111], K. Ganga[71], J. García-Bellido[98], E. Gaztanaga[88,87,112], F. Giacomini[16], G. Gozaliasl[113,62], A. Gregorio[103,12,13], A. Hall[33], H. Hildebrandt[114], J. Hjorth[115], J. J. E. Kajava[116,117], V. Kansal[118,119,120], D. Karagiannis[121,122], C. C. Kirkpatrick[59], L. Legrand[123], G. Libet[79], A. Loureiro[124,125], G. Maggio[12], M. Magliocchetti[44], F. Mannucci[126], R. Maoli[127,29], C. J. A. P. Martins[128,1], S. Matthew[33], L. Maurin[129], R. B. Metcalf[30,7], P. Monaco[103,12,13,14], C. Moretti[11,97,12,14,13], G. Morgante[7], Nicholas A. Walton[130], J. Odier[131], L. Patrizii[16], M. Pöntinen[62], V. Popa[80], C. Porciani[66], D. Potter[132], I. Risso[133], P.-F. Rocci[129], M. Sahlén[134], A. Schneider[132], M. Sereno[7,16], P. Simon[66], A. Spurio Mancini[135], C. Tao[60], G. Testera[20], R. Teyssier[136], S. Toft[56,137,138], S. Tosi[19,9,20], A. Troja[5,43], M. Tucci[42], C. Valieri[16], J. Valiviita[62,63], D. Vergani[7], and G. Verza[139,140]

*(Affiliations can be found after the references)*

## ABSTRACT

The *Euclid* Space Telescope will image about 14 000 deg$^2$ of the extragalactic sky at visible and near-infrared (NIR) wavelengths, providing a dataset of unprecedented size and richness that will facilitate a multitude of studies into the evolution of galaxies. Although spectroscopy will also be available for some of the galaxies, in the vast majority of cases the main source of information will come from broad-band images and data products thereof (i.e., photometry). Therefore, there is a pressing need to identify or develop scalable yet reliable methodologies to estimate the redshift and physical properties of galaxies using broad-band photometry from *Euclid*, optionally including ground-based optical photometry also. To address this need, we present a novel method developed as part of a 'data challenge' within the Euclid Collaboration, to estimate the redshift, stellar mass, star-formation rate, specific star-formation rate, $E(B-V)$, and age of galaxies, using mock *Euclid* and ground-based photometry. The main novelty of our property-estimation pipeline is its use of the `CatBoost` implementation of gradient-boosted regression-trees, together with chained regression and an intelligent, automatic optimization of the training data. The pipeline also includes a computationally-efficient method to estimate prediction uncertainties, and, in the absence of ground-truth labels, provides accurate predictions for metrics of model performance up to $z \sim 2$. We apply our pipeline to several datasets consisting of mock *Euclid* broad-band photometry and mock ground-based *ugriz* photometry, with the objective of evaluating the performance of our methodology for estimating the redshift and physical properties of galaxies detected in the Euclid Wide Survey. The statistical metrics of prediction residuals vary depending on which mock catalogue and filters are tested. Nonetheless, the quality of our photometric redshift and physical property estimates are highly competitive overall, validating our modeling approach. However, at $z \gtrsim 3.5$ the relative sparsity of galaxies resulted in unreliable redshift and physical property estimates, which we argue could be mitigated by building catalogues with better sampling of $z \gtrsim 3.5$ galaxies, or switching to the use of spectral energy distribution fitting in this regime. We also find that the inclusion of ground-based optical photometry significantly improves the quality of the property estimation, highlighting the importance of combining *Euclid* data with ancillary ground-based data from the Vera C. Rubin Observatory Legacy Survey of Space and Time (LSST), UNIONS, etc.

**Key words.** Galaxies: photometry – Galaxies: high-redshift – Galaxies: evolution – Galaxies: general

## 1. Introduction

Large-area observational surveys play an increasingly pivotal rôle in the adjacent fields of cosmology, astronomy, and astrophysics. By observing many millions, or even billions, of sources at high spatial resolution and with point-spread-function stability, such surveys aim to test and refine cosmological theory, while also generating extremely rich datasets enabling a multitude of extragalactic science questions to be potentially addressed [see, e.g., the Square Kilometer Array (SKA): Dewdney et al. 2009; the 4-metre Multi-Object Spectroscopic Telescope (4MOST): Guiglion et al. 2019; the Nancy Grace Roman Space Telescope (NGRST): Akeson et al. 2019; the Vera C. Rubin Observatory Legacy Survey of Space and Time (LSST): Ivezić et al. 2019; the Dark Energy Spectroscopic Instrument survey (DESI): Dey et al. 2019].

During the next several years and beyond, the *Euclid* Space Telescope will boost significantly our understanding of the evolution of galaxies across cosmic time. A $\sim 14\,000\,\mathrm{deg}^2$ area of the extragalactic sky will be imaged at visible and near-infrared (NIR) wavelengths, to a $5\,\sigma$ point-source depth of 26.2 mag[1] in the $I_\mathrm{E}$ ($R$+$I$+$Z$) filter of the Visible Instrument (VIS; Euclid Collaboration: Cropper et al. 2024), and 24.5 mag in the $Y_\mathrm{E}$, $J_\mathrm{E}$, and $H_\mathrm{E}$ filters (Euclid Collaboration: Scaramella et al. 2022; Euclid Collaboration: Schirmer et al. 2022) of the Near-Infrared Spectrometer and Photometer (NISP; Euclid Collaboration: Jahnke et al. 2024). Three additional fields with a combined area of 53 $\mathrm{deg}^2$ will be observed 2 magnitudes deeper, to a $5\,\sigma$ depth of 28.2 mag in the $I_\mathrm{E}$ band, and 26.5 mag in the $Y_\mathrm{E}$, $J_\mathrm{E}$, and $H_\mathrm{E}$ bands.

The *Euclid* surveys will provide multi-colour, broad-band imaging, allowing the detection of approximately 12 billion sources at $3\,\sigma$ significance or higher, and is also expected to yield spectroscopic redshifts for roughly 35 million galaxies (e.g. Laureijs et al. 2011; Euclid Collaboration: Mellier et al. 2024). Thus, *Euclid* observations are expected to make possible a diversity of unique extragalactic science, especially when combined with multi-wavelength observations from other large surveys, including the detection and study of very large samples of star-forming, passive, or active galaxies across cosmic time (see Euclid Collaboration: Mellier et al. 2024).

A crucial step towards extracting science from these data is the assignment of labels using parameters measured from images, to provide a characterisation of each galaxy (e.g., redshift, stellar mass, star-formation activity, the presence of nuclear activity, etc.). A widespread methodology is the use of software that compares spectral templates to an observed photometric spectral energy distribution (SED) or spectrum, deriving physical parameters from best-fitting templates (e.g., Arnouts et al. 1999; Bolzonella, Miralles & Pelló 2000; Cid Fernandes et al. 2005; Ilbert et al. 2006; da Cunha, Charlot & Elbaz 2008; Noll et al. 2009; Laigle et al. 2016; Gomes & Papaderos 2017; Carnall et al. 2018; Johnson et al. 2021; Pacifici et al. 2023). However, because the computation time typically scales linearly with the number of objects to be fitted, this family of methods can become computationally very expensive when applied to very large sets of data (i.e., $\gg 10^6$ objects).

Machine-learning methods offer an alternative (or complementary) approach that can be significantly more scalable than traditional template-fitting methods: Most of the computational cost is front-loaded in the model-training phase, with inference having only a marginal cost per object. Supervised learning is currently the most popular machine-learning paradigm for the classification of galaxies and for the estimation of their redshift and physical properties. In the supervised paradigm, the model-training process usually involves learning a function that aims to map observed values (e.g., magnitudes and colours) to labels (e.g., object class, redshift, etc.), using a statistical-learning algorithm such as a decision-tree ensemble (e.g., Breiman 2001) or an artificial neural network (e.g., McCulloch & Pitts 1943; Hinton 1989). Once trained, the model is then used for label inference at a relatively low computational cost (e.g., Hemmati et al. 2019). Potential limitations can include the need for a large amount of training data, biases, or issues with interpretability.

Helped by the availability of ready-to-use machine-learning methods in open-source packages such as `Scikit-Learn` (Pedregosa et al. 2011), there is now an exponentially-growing body of literature related to the application of supervised machine learning for source classification, and the estimation of the redshift and physical properties of galaxies. Among the most fundamental tasks is the classification of sources using broad-band photometry data, including the separation of sources into stars, quasars, and galaxies (e.g., Bai et al. 2019; Clarke et al. 2020; Cunha & Humphrey 2022), or the selection of specific classes of galaxy or quasars (e.g., Cavuoti et al. 2014; Signor et al. 2024; Euclid Collaboration: Humphrey et al. 2023; Cunha et al. 2024).

There has also been a multitude of studies in which deep-learning techniques are applied to the problem of automatically classifying galaxy images, with impressive results (e.g., Dieleman, Willett & Dambre 2015; Huertas-Company et al. 2015; Domínguez Sánchez et al. 2018; Tuccillo et al. 2018; Nolte et al. 2019; Bowles et al. 2021; Bretonnière et al. 2021; Li et al. 2022a), or for the identification and modeling of gravitational lenses (e.g., Petrillo et al. 2017; Gentile et al. 2023).

Another common use-case for supervised learning is the estimation of galaxy redshifts (e.g., Collister & Lahav 2004; Brescia et al. 2013; Cavuoti et al. 2017; Pasquet et al. 2019; Razim et al. 2021; Guarneri et al. 2021; Carvajal et al. 2021; Cunha & Humphrey 2022; Li et al. 2022b). Despite usually lacking the physical foundations of traditional template-fitting methods, it has been found that under some circumstances, supervised machine learning can outperform the traditional methods (Euclid Collaboration: Desprez et al. 2020). The reason why supervised machine learning sometimes outperforms traditional methods is primarily due to differences in inductive bias, including greater freedom in how observables are used. For instance, supervised learning algorithms may learn priors from the training data, can learn how to optimally weight observational inputs to obtain more accurate prediction outputs, and have the ability to recognise hidden relationships or physics that are not included in galaxy template recipes (see, e.g., Euclid Collaboration: Humphrey et al. 2023).

⋆ e-mail: `Andrew.Humphrey@astro.up.pt`

[1] We use AB magnitudes herein.

The estimation of physical properties of galaxies, such as stellar mass and star-formation rate, represents yet another attractive application for supervised learning (e.g., Ucci et al. 2018; Bonjean et al. 2019; Delli Veneri et al. 2019; Mucesh et al. 2021; Simet et al. 2021; Euclid Collaboration: Bisigello et al. 2023). This endeavour promises to be highly fruitful, facilitating the study of galaxy evolution across cosmic time with the enormous samples of galaxies that will soon become available from wide-area surveys such as those to be performed by *Rubin*/LSST and *Euclid*.

Beyond the purely supervised paradigm, there is a substantial number of extragalactic studies using unsupervised or semi-supervised machine-learning methods. For instance, Humphrey et al. (2023) recently demonstrated that the semi-supervised method known as 'pseudo-labeling' (Lee 2013) can be used to significantly improve some supervised machine-learning models by allowing the algorithm to also learn about the properties of the unlabeled (i.e., test) data. In addition, Cunha et al. (2024) presented a novel semi-supervised learning methodology for the identification of obscured quasars at high-redshift. Unsupervised methods, which generally do not make use of labels, have also been employed for a number of different tasks, including the separation of sources into statistically meaningful classes or clusters (e.g., Logan & Fotopoulou 2020), or the identification of rare or anomalous sources (e.g., Reis et al. 2018; Pruzhinskaya et al. 2019; Solarz et al. 2020).

A number of more exotic methods to augment supervised machine learning have also been explored, including active learning, where the model outputs help the user to improve the training data so as to improve model quality (e.g., Wolpert 1992), meta-learning, where a machine-learning algorithm learns about itself or other models (e.g., Zitlau et al. 2016; Euclid Collaboration: Humphrey et al. 2023), and hybrid approaches where results from traditional template-fitting methods are combined with machine-learning methods (e.g., Cavuoti et al. 2017; Fotopoulou & Paltani 2018).

In this study, we describe a novel supervised-learning methodology for the estimation of the redshift and physical properties of galaxies, using broad-band photometry measurements as input data. Although our work is focused on the application of this method to *Euclid*, LSST and UNIONS (Chambers et al. 2020) photometry, we emphasise that our methodology is data-agnostic and can be readily adapted and used with essentially any tabular dataset.

Our methodology aims to overcome a number of shortcomings in ML-based workflows for galaxy physical property estimation. In particular, our approach combines (i) the state-of-the-art `CatBoost` learning algorithm, (ii) an intelligent algorithm to optimise the composition of the input data, (iii) an attention mechanism that gives the learning algorithm awareness of multiple labels at once, and (iv) an efficient ML-based method to estimate prediction uncertainties. We emphasise that this study was performed in the context of a 'data challenge' within the Euclid Collaboration (see also Euclid Collaboration: Bisigello et al. 2023; Euclid Collaboration et al. 2024) and as such, its scope is limited to presenting our methodology and its results when applied to several mock *Euclid* galaxy catalogues. More detailed benchmarking and a comparison between different methods is presented in Euclid Collaboration et al. (2024).

This paper is structured as follows. In Sect. 2 we describe the rescaling of labels. Next, in Sect. 3, we define the different combinations of filters we use as test-cases. In Sect. 4 the datasets are described. The metrics we use to evaluate model quality are detailed in Sect. 5. The machine-learning pipeline is presented in Sect. 6. In Sect. 7 the results are described, and in Sect. 8 we present our conclusions.

## 2. Target label scalings

This study is principally concerned with the estimation of the redshift ($z$),[2] stellar mass ($M$), and star-formation rate (SFR) of galaxies. Before model training begins, most of the target labels are modified or rescaled to provide a distribution that is more straightforward for the learning algorithm to work with.

In the case of redshift, our pipeline adds the scalar value 1 to the redshifts prior to the model training. Experiments as part of this study, and our prior experience, indicate that using $1 + z$ generally gives superior results.

All but one of the other target labels are rescaled to have a logarithmic distribution, which our experiments and previous experience show generally improves model quality. The reference values[3] of $M$ are rescaled as

$$M_{\rm ref} = \log_{10}\left(\frac{\text{stellar mass}}{M_\odot}\right), \tag{1}$$

and the SFR is rescaled as

$$\text{SFR}_{\rm ref} = \log_{10}\left(\frac{\text{star-formation rate}}{M_\odot\,\text{yr}^{-1}}\right). \tag{2}$$

Similarly, the specific star-formation rate (sSFR) is rescaled as

$$\text{sSFR}_{\rm ref} = \log_{10}\left(\frac{\text{specific star-formation rate}}{\text{yr}^{-1}}\right). \tag{3}$$

The $M$ labels are rescaled as in Eq. (1), SFR labels are rescaled as in Eq. (2), and the sSFR labels rescaled as in Eq. (3).

Another label that is interesting to predict is the stellar age (hereinafter referred to simply as 'age'), defined as the time since the start of the first episode of star-formation. The age is rescaled as

$$\text{age}_{\rm ref} = \log_{10}\left(\frac{\text{stellar age}}{\text{yr}}\right). \tag{4}$$

All the quoted (or plotted) values of $M$, SFR, sSFR, or age have been rescaled as described above. However, the colour-excess $E(B-V)$ values do not require transformation since they are already logarithmic.

[2] We use the term 'redshift' and the symbol '$z$' interchangeably, with the aim of minimizing ambiguity with the $z$-band filter.

[3] Throughout this paper, the reference (or ground-truth) of a variable are denoted by the 'ref' subscript suffix, and the estimated (predicted) values are denoted by the 'est' subscript suffix.

## 3. Test-cases

In the interest of 'open science' and reproducibility, our initial test case makes use of a subset of the publicly-available COSMOS 2015 photometry catalogue of Laigle et al. (2016). This catalogue contains deep, multi-band photometry over the 2 deg$^2$ area of the COSMOS field, and provides high-quality photometric redshifts, $M$-esimates, and other physical properties or parameters; the authors used the spectral template-fitting code `LePhare` (Arnouts et al. 2007; Ilbert et al. 2006) to derive these properties, adopting a Chabrier initial mass function (Chabrier 2003). The COSMOS 2015 catalogue adopts a flat cosmology with dimensionless Hubble parameter $h = 0.7$, mass density $\Omega_{\rm m} = 0.3$, and cosmological constant $\Omega_\Lambda = 0.7$.

We use 3$''$ aperture photometry in the $u$, $B$, $V$, $r$, $i^+$, $z^+$, $Y$, $J$, $H$, $K_{\rm s}$ bands, corrected for Galactic extinction as prescribed in Laigle et al. (2016). We include only galaxies using the `TYPE=0` criterion, which excludes active galactic nuclei (AGN) and stars. We note that excluding AGN alters the bias of the sample, since galaxies in which the central supermassive black hole is undergoing significant accretion-driven growth are no longer present. We also exclude sources with photometric redshift values lower than 0 or higher than 9.9, to avoid unphysical redshift values. The selected galaxies also have good-quality photometry, with all sources having `FLAG_PETER` and `FLAG_HJMCC` equal to 0. To probe a generally similar region of magnitude-space as the Euclid Wide Survey, we use only galaxies with $H \leq 24$ mag, corresponding to an $H$-band signal-to-noise ratio (SNR) cut of $\sim 3.6$. The resulting catalogue contains 194 349 galaxies. To allow other teams to benchmark their methods against ours, we make this dataset available in the accompanying online material.

We also define several test-cases that represent expected real-world use-cases for *Euclid* photometry, with $\geq 3\,\sigma$ or $\geq 10\,\sigma$ detections, with or without ancillary ground-based photometry from, for example, LSST (Ivezić et al. 2019) or UNIONS (e.g., Chambers et al. 2020). In all cases, active galactic nuclei and sources with a detection in X-rays were excluded.

Thus, our test-cases are as follows:

- Case 0: COSMOS 2015 $u$, $B$, $V$, $r$, $i^+$, $z^+$, $Y$, $J$, $H$, $K_{\rm s}$ bands ($H \leq 24$ mag)
- Case 1: *Euclid* only ($\geq 3\,\sigma$ detections);
- Case 2: *Euclid* only ($\geq 10\,\sigma$ detections);
- Case 3: *Euclid* ($\geq 3\,\sigma$ detections) and *ugriz* bands (including non-detections);
- Case 4: *Euclid* ($\geq 10\,\sigma$ detections) and *ugriz* bands (including non-detections);

The number of galaxies ($N$) used for each combination of case and catalogue, and the main characteristics thereof, are shown in Table 2. In the interest of open science, the data used for Case 0 have been made available on Github, along with a version of our pipeline[4].

## 4. Mock *Euclid* galaxy catalogues

In Fig. 1, we show the distribution of galaxies as a function of $H_{\rm E}$ or redshift, for the four *Euclid* mock catalogues used in this study. The construction of the mock catalogues is described below. Note that in all catalogues, SFR and sSFR are instantaneous quantities.

[4] github.com/humphrey-and-the-machine/Euclid-chained-regression

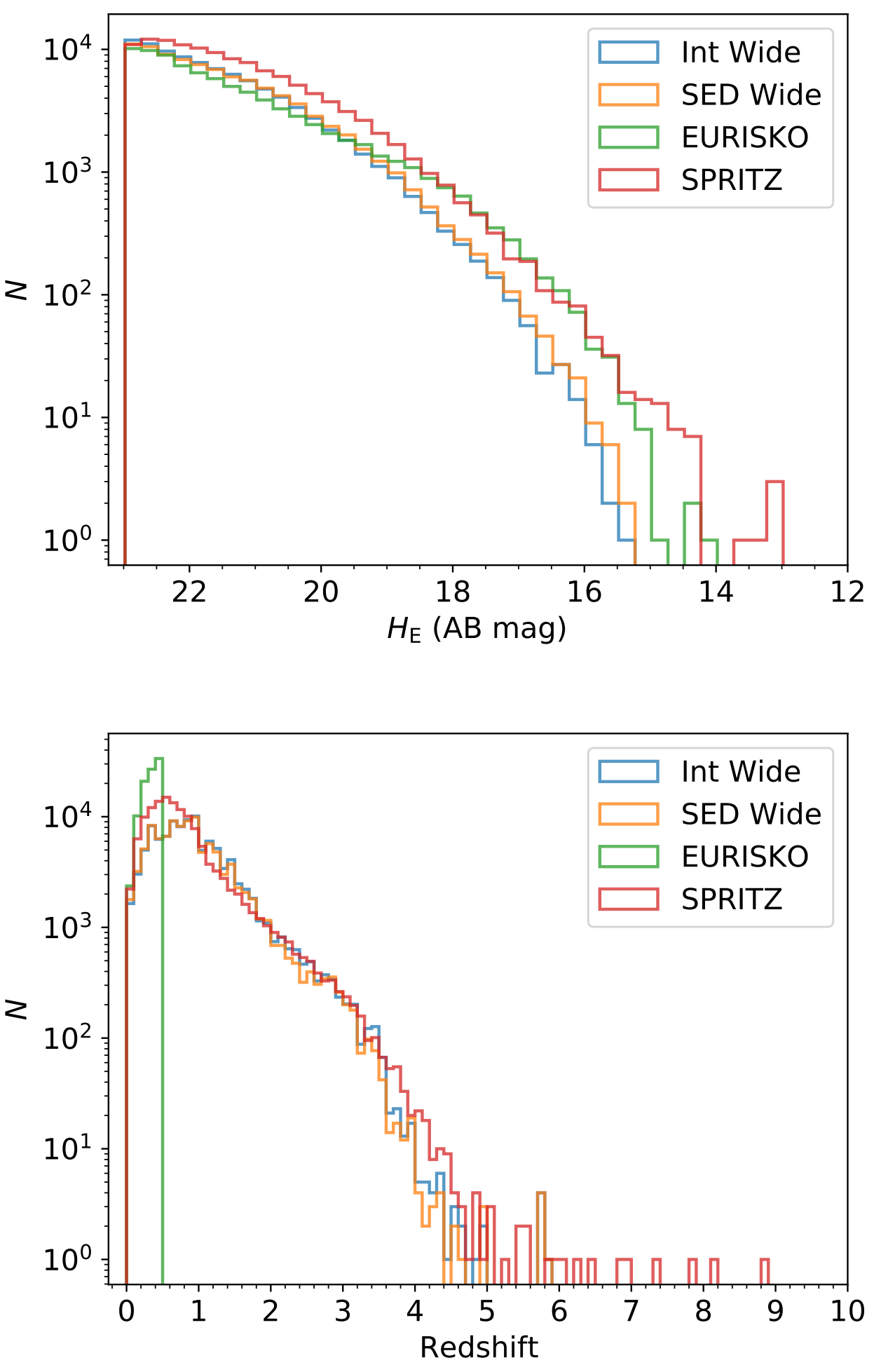

**Fig. 1.** Histograms of the number of sources as a function of $H_{\rm E}$ for the Int Wide, SED Wide, EURISKO, and SPRITZ mock *Euclid* catalogues (top), or the number of sources as a function of redshift (bottom). For consistency with the test cases described in Sect. 3, we include only sources that have a $\geq 3\,\sigma$ detection in the $Y_{\rm E}$, $J_{\rm E}$, and $H_{\rm E}$ filters. The histogram for COSMOS 2015 (Case 0; not shown) is similar to those of the Int Wide and SED Wide catalogues.

### 4.1. *Int Wide*

The Int Wide catalogue was produced by Bisigello et al. (2020) to simulate the Euclid Wide Survey (Euclid Collaboration: Scaramella et al. 2022), and is derived from the COSMOS2015 catalogue of Laigle et al. (2016). The Int Wide catalogue initially included the Canada-France Imaging Survey $u$ filter (CFIS/$u$) band and the *Euclid* $I_{\rm E}$, $Y_{\rm E}$, $J_{\rm E}$, $H_{\rm E}$ bands (Bisigello et al. 2020), and was later expanded to also include the *Rubin*/LSST *griz*, Wide-field Infrared Survey Explorer 3.4 and 4.6 $\mu$m (Wright et al. 2010) and 20 cm Very Large Array bands (Euclid Collaboration: Humphrey et al. 2023). The construction of the catalogue was described in detail by Bisigello et al. (2020) and Euclid Collaboration: Humphrey et al. (2023); here we provide a summary of the steps used in the construction.

- The COSMOS2015 multi-wavelength catalogue of Laigle et al. (2016) is the starting point.
- All sources that are labelled as stars or X-ray sources were removed, as were sources that were masked in optical broadbands, reducing the catalogue to 518 404 objects at $z \leq 6$.

- Next, a broken-line template from the ultraviolet to the infrared was produced for each source, by interpolation over the broad-band photometry.
- Finally, the template was convolved with the *Euclid* $I_{\rm E}$, $Y_{\rm E}$, $J_{\rm E}$, and $H_{\rm E}$ filters (Euclid Collaboration: Schirmer et al. 2022) to derive mock *Euclid* photometry.

Since the photometric errors are similar to (or larger than) those expected for the Euclid Wide Survey (Euclid Collaboration: Scaramella et al. 2022), it was not necessary to inject any artificial photometric scatter. It is important to note that although this catalogue is also based on the COSMOS2015 catalogue, the selection criteria differ from those used in Case 0 described in Sect. 3. This mock catalogue uses the cosmological parameter values $h = 0.7$, $\Omega_{\rm m} = 0.3$, and $\Omega_\Lambda = 0.7$ and the same Chabrier initial mass function (Chabrier 2003).

### 4.2. SED Wide

The SED Wide catalogue was also produced by Bisigello et al. (2020), using an alternative methodology to that described in Sect. 4.1. As before, objects labelled as X-ray sources or stars, and sources that were flagged as having been masked in optical broad-bands, were first removed. The spectral template-fitting code `LePhare` was then used to perform fitting of the COSMOS2015 photometry with a large set of Bruzual & Charlot (2003) templates. Redshifts were fixed at their COSMOS2015 values from Laigle et al. (2016). Metallicities of $Z_\odot$ or 0.4 $Z_\odot$ were considered, while star-formation histories with an e-folding timescale $\tau$ between 0.1 and 10 Gyr, and ages from 0.1 to 12 Gyr, were used. These ranges were chosen to strike a balance between having a manageable number of templates, and having physically reasonable coverage of the parameter space. The reddening law of Calzetti et al. (2000) was adopted, and 12 values of colour excess between 0 to 1 were considered. For each galaxy, the best template was identified via a $\chi^2$ minimisation. This template was then convolved with the *Euclid* filter transmission functions, to produce mock broad-band photometry. Finally, random (Gaussian) noise was added to this mock photometry, corresponding to the expected photometric errors in the Euclid Wide Survey (Euclid Collaboration: Scaramella et al. 2022). Ten copies of each source were produced, using different random noise realisations. It is important to note that the resulting mock photometry SED is a synthetic representation of the observed one, and for some sources the photometry or colours differ significantly from their observed values (see also Euclid Collaboration: Humphrey et al. 2023). This catalogue adopts the same cosmology as used in Sect. 4.1.

### 4.3. EURISKO

EURISKO (EUclid and *Rubin* photometry Inferred from SED fitting of Kids Observations) is a semi-empirical sample based on $\sim$ 122 500 galaxies with KiDS+ViKING photometry from Data Release 4 of the Kilo Degree Survey (KiDS-DR4) at $z <$ 0.5 (Kuijken et al. 2019).

To assemble the sample, we have extracted a random set of 10 KiDS tiles (1 deg$^2$ each, 5 in the north and 5 in the south caps) from KiDS-DR4 release, after removing masked regions, corresponding to a total effective area of $\sim$ 6.9 deg$^2$. The tiles are also in KiDS-DR3. The catalogues are publicly available.[5] We have extracted from the catalogues:

- The 9-band GaAP magnitudes ($u, g, r, i, Z, Y, J, H, K_{\rm s}$), which are in AB format, and already corrected for Galactic extinction (using the Schlafly & Finkbeiner 2011 prescription);
- photometric redshifts, determined using `BPZ` by the KiDS collaboration;
- the `FLUX_RADIUS`, used as an indicator of galaxy size, converted to arcsec using the OmegaCam pixel scale 0.2 arcsec/pix;
- The `2DOPHOT` star-galaxy separation, `SG2DPHOT`, which is equal to 0 for galaxies;
- the `MASK` parameter to select galaxies with the safest photometry, not affected e.g. by star halos.

The following selection criteria have been applied: a) `SG2DPHOT` = 0 to select galaxies; b) `MASK` = 0 to remove objects in masked regions; c) photometric redshift < 0.5 (the dataset was created to support studies of the low-$z$ Universe).

To create the mock *Euclid* and LSST magnitudes, we use `LePhare` to perform $\chi^2$ fitting between the stellar population synthesis theoretical models and KiDS data. With the redshift fixed at the value determined by the KiDS collaboration (see above), we fit the models to the data using the 9 GaAP bands (excluding for each galaxy the bands not available from the fit) and adopt Bruzual & Charlot (2003) synthetic models, assuming a Chabrier initial mass function (Chabrier 2003), implementing different metallicities in the range 0.2–2.5 $Z_\odot$, an exponential SFR with time duration $\tau$ from 0.1 to 30 Gyr and galaxy ages up to 13.5 Gyr. Internal extinction is accounted for using the Calzetti extinction curve and $E(B-V) = 0, 0.1, 0.2, 0.3, 0.4, 0.5$. Emission lines are added using the prescription provided in `LePhare`. A flat cosmology is adopted, with dimensionless Hubble constant $h = 0.7$, mass density parameter $\Omega_{\rm m} = 0.3$, and cosmological constant $\Omega_\Lambda = 0.7$. After running `LePhare`, and a best-fitted model is found, model magnitudes are obtained for *Euclid* and *Rubin*/LSST bands.

To determine realistic errors on the output magnitudes, we use

$$ {\rm d}f = \sqrt{{\rm d}f_{\rm bkg}^2 + {\rm d}f_{\rm obj}^2} = \frac{f_{\rm lim}}{\rm SNR}\frac{r}{r_{\rm ref}}\sqrt{1 + \frac{f}{f_{\rm sky}\pi r^2}}\,, \tag{5} $$

which depends on galaxy flux $f$, limiting flux $f_{\rm lim}$ (10 $\sigma$ detection limit), the related SNR, the sky surface brightness $f_{\rm sky}$, a typical galaxy radius $r$, and a reference value for it at the magnitude limit $r_{\rm ref}$. This corresponds to the contribution of the Poisson noise associated with the number of photons received from the background and from the source; rather than estimating it precisely from the detector properties, we instead rescale it to correspond to the median SNR at the limiting magnitude. For the value of $r$ we adopt the `FLUX_RADIUS`, assuming (for simplicity) that it is constant as a function of wavelength. For $r_{\rm ref}$ we adopt the value $0\farcs39$, which is the median value of galaxies in the KiDS $r$-band magnitude range 24.5–25.0. We use limiting magnitudes at 10 $\sigma$ (SNR = 10). The resulting errors are converted to magnitude errors using a standard error propagation as ${\rm d}m = 2.5\,{\rm d}f\,/\,[\,\ln(10)\,f\,]$, an approximation that results in errors that are symmetric in magnitudes.

### 4.4. SPRITZ

The Spectro-Photometric Realisations of Infrared-selected Targets at all-$z$ (SPRITZ; Bisigello et al. 2021) was derived using the IR luminosity functions observed by Herschel up to $z$ ∼3.5

[5] kids.strw.leidenuniv.nl/.../KiDS_Synoptic_Table_Catalogview.php

(Gruppioni et al. 2013), the $K$-band luminosity function of elliptical galaxies (Arnouts et al. 2007; Cirasuolo et al. 2007; Beare et al. 2019), and the galaxy stellar-mass function of dwarf-irregular galaxies (Huertas-Company et al. 2016; Moffett et al. 2016). The simulation contains star-forming galaxies (i.e. spirals, starbursts, and dwarfs), passive galaxies, AGN, and composite systems where an AGN is present but is not the dominant source of power.

A set of SED models (Polletta et al. 2007; Rieke et al. 2009; Gruppioni et al. 2010; Bianchi et al. 2018), considering a Chabrier initial mass function (Chabrier 2003), was assigned to each simulated galaxy, and photometric fluxes expected in the *Euclid* filters were then extracted. Photometric (Gaussian) noise consistent with that expected in the Euclid Wide Survey (Euclid Collaboration: Scaramella et al. 2022) was added. Physical properties (e.g. $M$, SFR) were then assigned, considering theoretical or empirical relations, or directly from the SED assigned to each simulated galaxy.

In the construction of this mock catalogue, Bisigello et al. (2021) adopted a ΛCDM cosmology with a dimensionless Hubble parameter $h = 0.7$, a mass density $\Omega_\mathrm{m} = 0.27$, and a cosmological constant $\Omega_\Lambda = 0.73$.

Overall, SPRITZ is consistent with a large set of observations, including luminosity functions and number counts from X-ray to radio, the global galaxy stellar-mass function, and the SFR vs. stellar-mass plane. See Bisigello et al. (2021) for more details on the simulation and for additional comparison with observations. Before making use of the SPRITZ Euclid Wide Survey mock catalogue, we remove galaxies containing an AGN (i.e., AGN objects and composite objects). Finally, we randomly under-sampled the SPRITZ catalogue to reduce the number of sources to a manageable size ($\sim 300\,000$ sources).

## 5. Metrics of model quality

The metrics we use to quantify the quality of our redshift and physical property estimates are detailed below. In the case of redshift, the metric formulae require a division by $1 + z$ to transform the residuals from linear to relative scale. For the other properties, such a transformation is not necessary, since they are already logarithmic. Unless otherwise stated, the statistical metrics are calculated over all galaxies in the test set, with all galaxies therein being weighted equally.

### 5.1. Redshift metrics

To quantify the degree to which our redshift estimations are in error, we adopt the normalised median absolute deviation (NMAD). This metric includes scaling factors such that the result is approximately equivalent to the standard relative deviation, with a reduced impact from extremely-outlying errors. We calculate the NMAD as

$$\mathrm{NMAD} = 1.48\,\mathrm{median}\left(\frac{|z_\mathrm{est} - z_\mathrm{ref}|}{1 + z_\mathrm{ref}}\right), \tag{6}$$

where $z_\mathrm{est}$ is the estimated redshift, and $z_\mathrm{ref}$ is the 'ground-truth' reference redshift value. The NMAD is broadly equivalent to the standard deviation; smaller values of this metric indicate higher-quality redshift predictions.

In addition, we define the fraction of catastrophic outliers ($f_\mathrm{out}$; see, e.g., Hildebrandt et al. 2010) using the criterion

$$\frac{|z_\mathrm{est} - z_\mathrm{ref}|}{1 + z_\mathrm{ref}} > 0.15. \tag{7}$$

We also calculate the overall bias in the redshift estimations as

$$\mathrm{bias} = \mathrm{median}\left(\frac{z_\mathrm{est} - z_\mathrm{ref}}{1 + z_\mathrm{ref}}\right), \tag{8}$$

where values closer to zero are better.

### 5.2. Physical parameter metrics

For the physical property estimates, we calculate NMAD, $f_\mathrm{out}$, and the bias using formulae that differ slightly to those in Sect. 5.1. In this case, we calculate NMAD as

$$\mathrm{NMAD} = 1.48\,\mathrm{median}\,|y_\mathrm{est} - y_\mathrm{ref}|, \tag{9}$$

where $y_\mathrm{est}$ is the estimated value of the physical property, and $y_\mathrm{ref}$ is its 'ground-truth' value.

For physical properties, we consider a prediction to be an outlier if it differs from the true value by a factor of 2 or more (i.e., 0.3 dex; see also Euclid Collaboration: Bisigello et al. 2023). Thus, since the physical conditions are in log scale, $f_\mathrm{out}$ is calculated as

$$|y_\mathrm{est} - y_\mathrm{ref}| > 0.3. \tag{10}$$

We calculate the bias in the physical property estimates as

$$\mathrm{bias} = \mathrm{median}\,(y_\mathrm{est} - y_\mathrm{ref}). \tag{11}$$

In addition, we calculate the mean absolute error of our physical property estimations as

$$\mathrm{MAE} = \frac{\sum |y_\mathrm{est} - y_\mathrm{ref}|}{n}, \tag{12}$$

where $n$ is the number of samples. Smaller values of MAE indicate smaller errors, on average.

Finally, we also calculate the coefficient of determination, $R^2$, as

$$R^2 = \frac{\sum |y_\mathrm{est} - y_\mathrm{ref}|}{\sum |y_\mathrm{est} - \bar{y}_\mathrm{ref}|}, \tag{13}$$

where $\bar{y}_\mathrm{ref}$ is the mean value of $y_\mathrm{ref}$. A higher value of $R^2$ indicates a higher-quality model, with a maximum possible value of 1.

## 6. The property-estimation pipeline

### 6.1. Data preprocessing

The preprocessing steps that are performed to prepare the data for model training are described below.

#### 6.1.1. Broad-band colours

Broad-band magnitudes form the starting basis of the features used for training the machine-learning models. Even though these magnitudes contain information on the spectral energy distribution of a galaxy, the task of the learning algorithm can be made simpler by also including broad-band colours. This strategy is backed-up by experiments we conducted, where removal of some colours, or using only the magnitudes, resulted in lower-performing models (requiring more iterations or producing lower-quality predictions). Thus, we compute all possible broad-band colour (unique) permutations, which are included as features along with the magnitude values. In the case where one or both magnitudes in a colour are missing, that colour is flagged as missing. See Sect. 6.2 for further details about this issue.

### 6.2. Missing data imputation strategy

Since real survey data will contain samples with missing values, due to non-detections or other circumstances, it is imperative that any methodology to estimate galaxy physical properties is able to work with missing data. This allows for larger and richer samples, and potentially higher-quality models since non-detections often carry information about the redshift and properties of those galaxies (e.g., Steidel et al. 1996). Our missing value imputation approach follows that of Euclid Collaboration: Humphrey et al. (2023), who replaced missing values with a 'magic value' of −99.9, under the premise that decision-tree ensembles such as the one used herein will use the presence of missing values to perform splits where useful. Although our pipeline has the capability to impute different values to denote different origins of the missing values (i.e., not observed, masked, or not detected), in the interest of simplicity we herein impute a only a single magic value. In a future study, we will explore more complex methodologies for flagging missing photometry, with the objective of providing the learning algorithm with a more direct and granular representation of the nature of missing photometry values.

### 6.3. Additional preprocessing steps

The dataset is split randomly into training and test sets, with a ratio of 2:1. This ratio, although somewhat arbitrary, was chosen to obtain what we expect to be a reasonable balance between having a large training sample (to train stronger models), and a test set that is large enough for the metrics of model performance to be representative of the overall dataset. A classical validation set is not needed with our methodology, since our pipeline does not need to perform hyperparameter optimization.

The training and test sets have essentially identical depths in all bands, since they are drawn from the same mock catalogue. Transfer learning, where significantly different datasets are used for training and inference, is beyond the scope of this study, and is deferred to a possible future publication.

The features are standardized by subtracting the mean value and dividing by the standard deviation, where both statistics are calculated in the training set only. Missing values are ignored during this process, and thus are propagated to the input datasets unchanged.

### 6.4. The learning algorithm

Gradient-boosting tree methods (see Friedman 2001) combine multiple weak models, typically single-tree models, to build a stronger prediction model. In a nutshell, this class of algorithm trains a series of weak models on top of each other, where at each iteration a new weak model is trained to predict the error from the previous iteration, and this new model is combined with the previous model to reduce the error. Over the course of this procedure, a strong model is built.

`CatBoost`[6] is a state-of-the-art gradient-boosting tree method, which contains a number of relevant innovations, including the use of 'ordered boosting' to overcome overfitting, and 'oblivious trees' to improve speed and provide additional regularization. `CatBoost` was selected for this study because it was, arguably, the most advanced gradient-boosting tree method to be publicly available at the time.

[6] https://catboost.ai; `version 0.26`

**Table 1.** Fixed `CatBoostRegressor` hyperparameters

| Hyperparameter | Simple model | Complex model |
|---|---|---|
| `n_estimators` | 500 | 2000 |
| `max_depth` | 4 | 10 |

#### 6.4.1. `CatBoostRegressor` hyperparameters

In this study, our `CatBoostRegressor` models are instantiated with one of two sets of hyperparameters. The 'simple model' is a light-weight model that requires relatively few resources to train. It is used within our pipeline when the compromise between speed of training and model performance needs to favour the former. For instance, the simple model is used in the re-weighting procedure (Sect. 6.4.2), and for various checks or tests where a quick result is needed and maximal model performance is not required.

The 'complex model', on the other hand, uses higher values for the parameters `n_estimators` and `max_depth`, to maximise model quality. The values of these hyperparameters are listed in Table 1. All other hyperparameters are left unspecified, which allows the `CatBoostRegressor` instance to dynamically select or change their values using internal heuristics, adapting to the properties of the training set (Prokhorenkova et al. 2018).

From the available objective (loss) functions, we select the one that is most similar to the NMAD formula used for a particular label: For redshift, we use the mean absolute percentage error, and for other properties we use the mean absolute error objective function.

We emphasise that operation of our pipeline is agnostic with respect to the physical assumptions, such as the adopted initial mass function or the cosmology, and it is neither possible nor relevant to impose such assumptions thereupon. For instance, in the event that a different cosmology is adopted, causing the label values to be differently scaled, our pipeline will simply learn a different mapping between the input features and the labels.

#### 6.4.2. Re-weighting attention mechanism

The `CatBoostRegressor` algorithm allows the user to specify the weight for each training example, such that a training example can be made relatively more important (or less so) in the model training process. A higher weight for an example (i.e. a galaxy or galaxy subset) results in it having a greater importance in the model training. Our objective here is for the pipeline to learn which subsets of the training data are more (or less) valuable for the model training. This approach can be viewed as analogous to 'attention' mechanisms used in some deep-learning architectures (e.g., Vaswani et al. 2017).

Prior to training the model, weights for different subsets of the training set are optimised on a per-label basis, using a grid-search. Specifically, the training data are first divided into multiple bins in label-space, and the default weight of 1 is initially assigned to all bins. Next, the bins and the possible weight-values are iterated over, with a simple model being trained at each of these iteration. The performance of these models is evaluated using the relevant NMAD formula and cross-validation, and the weight-values that result in the lowest NMAD score are adopted. In the case where the NMAD is not affected by the choice of weight-value, the default weight of 1 is kept.

For the results presented herein, this re-weighting process is performed only for the redshift, $M$, and SFR labels. When properties other than these are modelled/predicted, the weights determined for redshift are adopted by default.

Compared to the case where the training examples are all weighted equally, the re-weighting procedure typically gives an improvement in the redshift NMAD score of ∼ 10 %, with the physical property estimates also usually receiving a significant improvement in their NMAD scores. These results highlight the usefulness of optimising the composition (weighting) of training data for a given generalisation task, and highlights the fact that a less representative training distribution may allow a stronger model to be trained (see, e.g., Euclid Collaboration: Bisigello et al. 2023).

### 6.4.3. Model training: Chained regression

Our pipeline applies the 'chained regression' methodology (see, e.g., Read et al. 2011; Cunha & Humphrey 2022) to the problem of predicting several scalar labels that exhibit significant covariance. In practical terms, the idea is to allow the learning algorithm to discover the covariance between the labels by iteratively predicting each label, with knowledge of its previous predictions of all the labels.

Our implementation of chained regression performs the following steps, which are summarized in Fig. 2. First, the training data is split into two folds of equal size, to allow out-of-fold (OOF) predictions to be made for the entire training set, without the risk of overfitting that is often present when a model is trained and predicts on the same examples. Next, for each of the two folds, a regression model is trained to predict one label, using the training data (the colours and magnitudes) as input. The model trained on one of the folds is used to predict OOF labels for the other fold, and vice versa. The OOF predictions are then appended as a new feature in the training. This is repeated sequentially for each label that is to be predicted. This constitutes one iteration of our chained regression pipeline. The second iteration starts again with the first label, this time using the training data with the previous OOF predictions as input. The new OOF predictions are appended as new features. In this way, each model that is trained has an awareness of previous label predictions. The procedure is repeated for the desired number of iterations, or until convergence is observed. Here, we find that four iterations is sufficient for convergence, which we define as detecting no significant additional improvement in the NMAD metric.

The final result of the model training is a regressor chain: a series of individual regression models that must be applied in the order in which they were trained. Predictions on unseen (test) data are made by applying the model chain to the test data. Due to the two-fold model training scheme we employ, there are two models, and thus two sets of predictions at each step in the regression chain; the two predictions are averaged to obtain a single prediction.

## 6.5. Estimating confidence intervals

### 6.5.1. Modeling prediction errors

In addition to point-estimates for redshift and the physical properties, it is also important to estimate confidence intervals for each prediction. For the properties estimated by the pipeline, uncertainties corresponding to the 68 % confidence interval are estimated by modeling the residuals between the predicted true labels (i.e., $|y_{\rm est} - y_{\rm ref}|$).

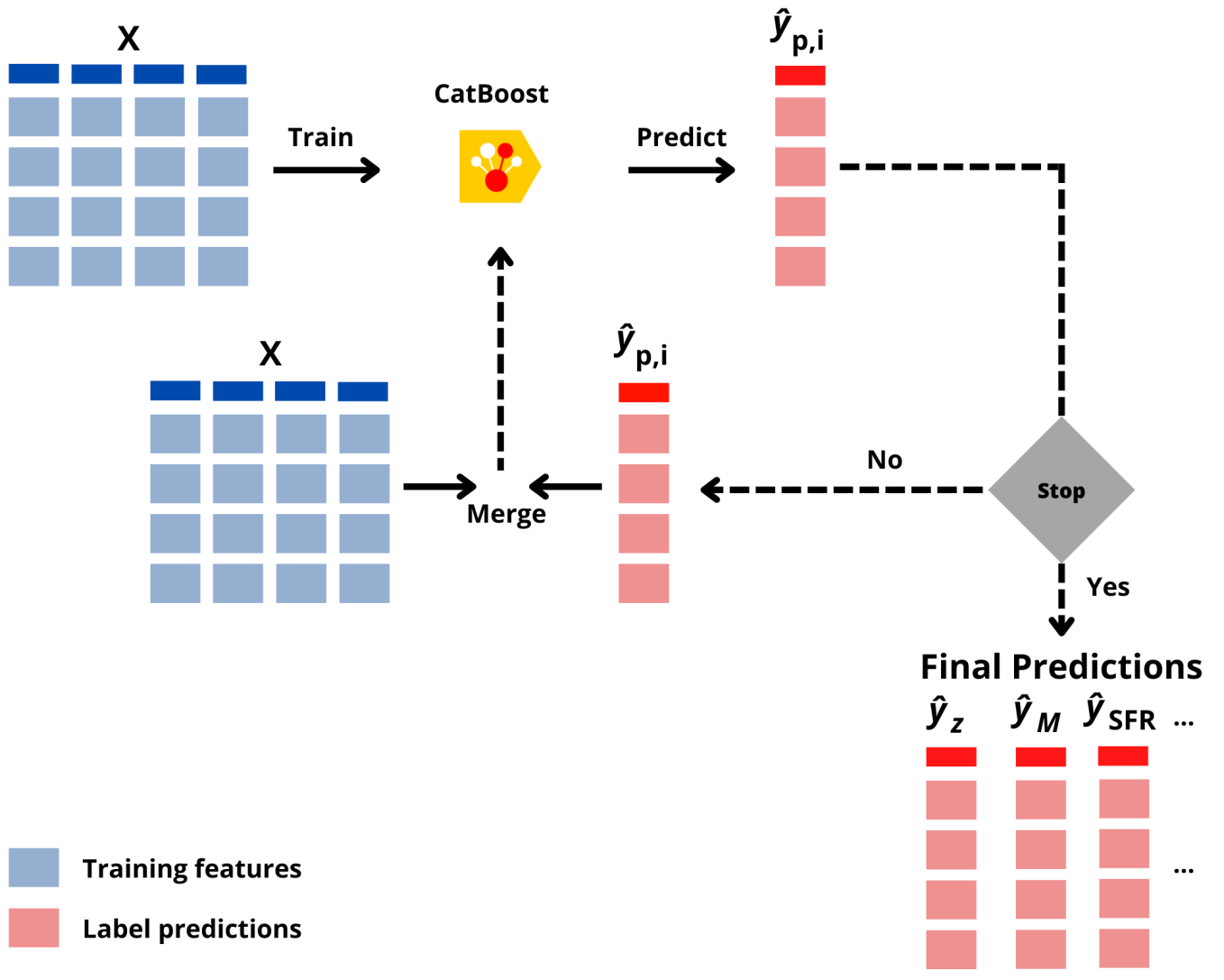

**Fig. 2.** Flow diagram summarizing the main steps in our chained regression implementation. In the first step, a `CatBoostRegressor` model is trained using the training data features X and training data labels $y$ (not shown) for one of the galaxy properties as inputs. The resulting model then provides predictions $\hat{y_{p,i}}$ for this galaxy property, both for the test set and the training set. These predictions are merged into to the training and test datasets as a new feature. This process is continued until each property has been predicted the required number of times, at which point the loop is stopped and the final predictions for each property are obtained.

We train a `CatBoostRegressor` 'simple model' that aims to directly predict the uncertainty in the individual redshift or physical property estimates. For this task, the training data comprises the training data used previously in Sect. 6.4, including the predicted values of redshift and physical conditions. In this case, the target labels are generated by subtracting the ground truth value from the predicted value of redshift or the physical properties. Although the model is trained to attempt to predict the residuals, its output predictions are essentially equivalent to the typical residual for each object, since the object-to-object randomness in the residuals cannot be predicted by the model. Due to the nature of this task, the Poisson objective function was used.

In Fig. A.1 we show the distribution of residuals with respect to the predicted 68 % confidence interval, when predicting redshift, $M$ or SFR, using the Int Wide catalogue with the Case 4 configuration. This figure confirms that the predicted uncertainty values are consistent with the measured 68 % uncertainties.

### 6.5.2. Estimating pipeline performance on unlabeled data

Our pipeline also estimates the quality of its predictions on unlabeled data, using the results of the uncertainty modeling described above (Sect. 6.5.1), with the assumption that the true errors (i.e., $|y_{\rm est} - y_{\rm ref}|$) are equal to the estimated errors. This is analogous to the 'confidence-based performance estimation' method applied to binary classification by Humphrey et al. (2022). Figure A.2 shows results from testing the performance of our error estimation method in different redshift bins. For redshift, the NMAD metric was estimated as

$$\mathrm{NMAD_{est}} = \mathrm{median}\left(\frac{\Delta z_{\rm est}}{1 + z_{\rm est}}\right), \tag{14}$$

where $\Delta z_{\rm est}$ is the predicted 68 % uncertainty of $z_{\rm est}$. Similarly, the NMAD metric was estimated for the physical properties as

$$\mathrm{NMAD_{est}} = \mathrm{median}\ (\ \Delta y_{\rm est}\ )\ , \quad (15)$$

where $\Delta y_{\rm est}$ is the predicted 68 % uncertainty of the estimated physical property value $y_{\rm est}$.

We use two different binning strategies. The first corresponds to the case where the ground truth is available, and thus the sources are binned by redshift using $z_{\rm ref}$. In the second method, the binning is performed using $z_{\rm est}$, and represents the 'real-world' case where the ground-truth labels are not available. Nevertheless, the results are similar when using either of the two binning methods.

From Fig. A.2, we note that in the $0 \le z \lesssim 2.5$ range, the values of $\mathrm{NMAD_{est}}$ are very similar to the measured values of NMAD, for the physical properties $M$ and SFR. At $z \gtrsim 2.5$, the measured NMAD increases much more rapidly with $z$ than does $\mathrm{NMAD_{est}}$. In the case of redshift, the $\mathrm{NMAD_{est}}$ is consistent with the measured NMAD only up to $z \sim 1$.

The cause of the underestimation of NMAD at high redshift is likely due to the relative sparsity of high-redshift sources in the training set, which makes it more challenging to learn the mapping between the broad-band SED and the target properties.

### 6.6. Computational efficiency

Among the well-known benefits of many machine-learning methods is their computational efficiency compared to that of some traditional SED-fitting methods. To provide some context about the relatively minimal computing resources that are required to run our pipeline, we have timed its execution on a mid-range laptop with a quad-core Intel i5-8350U CPU and 16 Gigabytes of RAM, running an Ubuntu Linux operating system. The total time required to perform all the steps in our pipeline, training on 71 015 randomly chosen examples from the Int Wide catalogue, using 4 iterations of chained regression, and six labels (redshift, SFR, sSFR, $M$, age, $E(B-V)$), is approximately 48 min for Case 1 (*Euclid* photometry and colours only) or 1 h 52 min for Case 3 (*Euclid* and *ugriz*). Once trained, the inference (prediction) of the labels is extremely fast, returning predictions for all six labels at a rate of $\sim 1.2 \times 10^{-4}$ s per galaxy, or $\sim 30$ h per billion galaxies. Our pipeline scales well with larger datasets and is set up to leverage power high-performance computing.

## 7. Results

### 7.1. Metric averaging methodology

It is crucial to ensure the metrics of model quality that we quote are representative, and not significantly influenced by a fortuitous (or unlucky) train-test split. Thus, the metric values are averaged over several runs, using a different random seed for the train-test splitting each time, to ensure the results are representative. The number of runs per Case ranged between 5 and 10, depending on the number of galaxies in the training dataset. As a general rule, having more galaxies resulted in a longer model training time, but a smaller variance in the metrics between runs.

The typical uncertainty on the average values of the metrics varies between the different Cases, and between the different metrics, but is usually smaller than 10 % of the metric value. In Cases where the number of galaxies is highest (e.g., Case 0), the variance between runs is negligible.

### 7.2. Case 0: Proof of concept

The results from applying our pipeline to the Case 0 (COSMOS) dataset are shown in Tables 3, where each table presents results from predicting redshift, $M$, SFR, sSFR, $E(B-V)$, or age. In Fig. 4, we plot the estimated properties versus their reference values (upper row), and plot the distribution of residuals (lower row).

In Table 4 we illustrate the improvement achieved using our chained regression approach for Case 0, compared to the case where each label is predicted using a single regression model. In Fig. 3 we show how the NMAD and $f_{\rm out}$ metrics for redshift, $M$, and SFR improve during four iterations of our pipeline. The results shown in this figure are the final results from the pipeline, for a single train-test split, and thus there may be small differences when compared to the averaged values shown in Table 3. Between the first and second iteration, there is a steep improvement in these metrics; the improvement continues more gently until the third or fourth iteration, after which we observe only a marginal improvement, or none. The size of the improvement varies from property to property, ranging between $\sim 5$ and $\sim 20$ per cent, with the redshift predictions showing a notably large improvement ($\sim$ 15–20 per cent). These results confirm our hypothesis that predicting several properties simultaneously in a chained-regression approach can lead to more reliable predictions for each one.

The improvements come from two main effects. First, by having an awareness of the previous prediction(s) of a label, the subsequent attempts to model the mapping between the features and this label can be more efficient, allowing the learning algorithm to spend less time on examples that are already well modeled, and more time on those examples that are not yet well modeled. In addition, some labels become less challenging to model when the learning algorithm has an awareness of the predicted values of other labels (e.g., having redshift estimates can facilitate more accurate estimation of $M$, and so on).

The metrics obtained for each of the properties are competitive compared to other results in the literature, for similar datasets (e.g., Fotopoulou & Paltani 2018; Euclid Collaboration: Desprez et al. 2020; Cunha & Humphrey 2022; Euclid Collaboration: Bisigello et al. 2023; Euclid Collaboration et al. 2024). For instance, Euclid Collaboration: Bisigello et al. (2023) reported NMAD($z$) $\sim$ 0.006–0.05, NMAD($M$) $\sim$ 0.04–0.2, and NMAD(SFR) $\sim$ 0.3–0.9, with which our metric values for these quantities overlap. It is particularly noteworthy that our redshift predictions are characterised by relatively low values for NMAD, outlier fraction, and bias. However, comparison between the results of different studies in the literature is fraught with complications, primarily due to the fact that different studies almost always adopt their own, somewhat different, datasets. Thus, we are unable to draw strong conclusions when comparing our results with those of previous studies.

We also remark on the special case of the problem of estimating the colour excess parameter $E(B-V)$. The fact that the $E(B-V)$ labels are quantized with steps of 0.1 means, clearly, that this label in particular contains significant noise (typical error $\sim 0.025$). Thus, it is likely that differences between the label and predicted values are at least partly due to errors in the label values, and thus the metric values for our $E(B-V)$ predictions likely understate the performance of our methodology. Furthermore, the fact that our models predict continuous (rather than quantized) values means that our predictions for $E(B-V)$ could potentially be closer to the actual ground truth than the original, quantized (noisy) labels.

**Table 2.** Overview of test Cases and catalogues, including photometric bands, selection cutoff, and the number of galaxies used.

| Catalogue | Case | Bands | Selection cutoff | $N$ |
|---|---|---|---|---|
| COSMOS | 0 | $uBVri^{+}z^{+}YJHK_{s}$ | $H \leq 24$ mag | 194 349 |
| Int Wide | 1 | *Euclid* only | $3\,\sigma$ (*Euclid*) | 105 993 |
| Int Wide | 2 | *Euclid* only | $10\,\sigma$ (*Euclid*) | 71 871 |
| Int Wide | 3 | *Euclid* + *ugriz* | $3\,\sigma$ (*Euclid*) | 105 993 |
| Int Wide | 4 | *Euclid* + *ugriz* | $10\,\sigma$ (*Euclid*) | 71 871 |
| SED Wide | 1 | *Euclid* only | $3\,\sigma$ (*Euclid*) | 103 422 |
| SED Wide | 2 | *Euclid* only | $10\,\sigma$ (*Euclid*) | 69 847 |
| SED Wide | 3 | *Euclid* + *ugriz* | $3\,\sigma$ (*Euclid*) | 103 467 |
| SED Wide | 4 | *Euclid* + *ugriz* | $10\,\sigma$ (*Euclid*) | 69 825 |
| EURISKO | 1 | *Euclid* only | $3\,\sigma$ (*Euclid*) | 93 827 |
| EURISKO | 2 | *Euclid* only | $10\,\sigma$ (*Euclid*) | 69 338 |
| EURISKO | 3 | *Euclid* + *ugriz* | $3\,\sigma$ (*Euclid*) | 93 827 |
| EURISKO | 4 | *Euclid* + *ugriz* | $10\,\sigma$ (*Euclid*) | 69 338 |
| SPRITZ | 1 | *Euclid* only | $3\,\sigma$ (*Euclid*) | 299 103 |
| SPRITZ | 2 | *Euclid* only | $10\,\sigma$ (*Euclid*) | 200 309 |
| SPRITZ | 3 | *Euclid* + *ugriz* | $3\,\sigma$ (*Euclid*) | 299 103 |
| SPRITZ | 4 | *Euclid* + *ugriz* | $10\,\sigma$ (*Euclid*) | 200 309 |

### 7.3. *Euclid* mock catalogues

In Figs. 5–9 and Fig. A.3 we plot the results from applying our pipeline to the mock *Euclid* datasets described in Sect. 3. The results are also listed in Table 3. As a general result, we find that the metrics vary between the different mock *Euclid* datasets and data configuration Cases. Unsurprisingly, including optical broad-band photometry (Case 3 & 4) usually provides a substantial improvement in model quality, compared to when only *Euclid* photometry is used (Case 1 & 2; see, e.g., Fig. 9). Furthermore, raising the minimum SNR cutoff from 3 to 10 also often gives a significant improvement. In other words, the NMAD,$f_{\rm out}$, and MAE metrics generally decrease, and $R^2$ generally increases, from Case 1 through 4. For the Int Wide, SED Wide and EURISKO catalogues, there is usually a large step-change in these metrics between Case 2 and Case 3, driven by the inclusion of the optical bands in Cases 3 and 4. For the SPRITZ catalogue, the metrics evolve more smoothly across the Cases.

In some cases, horizontal structure is visible in the density plot (e.g., Fig. 3), indicating a degeneracy such that the model has difficulty in choosing between several potential parameter values; this problem is diminished with the inclusion of optical photometry, and the use of the SNR = 10 cutoff.

Even when using an identical set of filters and the same minimum SNR cut-off, the quality of our redshift and physical property estimates varies between the catalogues, often dramatically so. For example, for a given Case the metrics we obtain using the EURISKO catalogue are vastly superior to those obtained for any of the other catalogues. For EURISKO, the values we obtain for the NMAD, MAE, and $f_{\rm out}$ metrics are typically a factor of $\sim 2$ smaller than those obtained, for a given Case, using the other catalogues. This is at least partly due to the fact that EURISKO contains a restricted redshift range ($0 < z < 0.5$), which simplifies substantially the learning problem. For instance, the potential for redshift and colour degeneracies to confuse the learning algorithm is greatly reduced, compared to catalogues that do not have a maximum redshift cut-off.

For the other catalogues, where the formal redshift cut-off is at $z = 6$, there are still significant differences in the various metrics. In the cases of the redshift, SFR, and sSFR predictions, we obtained better metric scores for the SPRITZ catalogue than for Int Wide or SED Wide. However, the reverse is true in the case of the $M$ predictions.

We find that the metric scores obtained with the Int Wide catalogue are similar to, or significantly better than, those obtained with the SED Wide catalogue. In particular, the metrics for $M$, and (for cases 3 and 4) the metrics for sSFR, $E(B-V)$ and age are significantly better for Int Wide than for SED Wide. This may be due to the fact the SED Wide catalogue contains somewhat simplified energy distributions, potentially erasing complex or unknown spectral features that are useful for estimating galaxy properties, making the regression problem more difficult. On the other hand, it is also possible that the labels of the Int Wide catalogue are slightly easier to predict, since they are predictions from another code (`LePhare` in this case) instead of being 'ground-truth' labels, and thus are likely contain simplifying biases.

Although we have tested the redshift range $0 \leq z \leq 6$ for all catalogues (except EURISKO, which is resticted to $z \leq 0.5$), we emphasise that our redshift predictions become rather unreliable at $z \gtrsim 3.5$. This is likely due to the sparsity of examples above this redshift range in the training data, making it challenging for the learning algorithm to learn how to reliably map the photometry and colour information to the redshift label. A knock-on effect of this is that the estimates of the other, physical properties are likely to be unreliable for galaxies at $z \gtrsim 3.5$.

In Fig. A.4 we illustrate how the NMAD metric varies with redshift, using results from a single model run that used the Case 4 data configuration with the Int Wide catalogue. The NMAD metric is generally at its lowest at $z \sim 1$, showing a gradual increase towards higher redshifts. In some cases, NMAD also shows a significant increase towards lower redshifts ($M$, SFR, sSFR, $E(B-V)$).

Overall, we find a substantial dispersion in metrics of model quality across the range of mock *Euclid* catalogues considered herein, with a strong dependence on whether *Euclid* photometry is used alone or with ancillary-optical photometry, and the way in which the mock catalogue is constructed. As such, we argue that using a single mock catalogue to simulate the performance of a method on real *Euclid* data is potentially risky. Furthermore, we argue that it is not necessarily a simple task to select the 'best' mock catalogue to forecast the model performance on *Euclid* data: paradoxically, one may choose between a dataset with fully realistic spectral shapes, but with biased labels, or a dataset with simplified spectral shapes and real 'ground-truth' labels, but obtaining the best of both worlds (i.e., realistic SEDs and 'ground-truth' labels) is not trivial.

Finally, we emphasise that the reported performance of some of the models may be optimistic. In the case of the Int Wide and Case 0 (COSMOS2015) catalogues, the labels we use to assess model performance are those derived from the SED-fitting of Laigle et al. (2016), which are not strictly 'ground-truth' values, and which have random or systematic errors with respect to the actual ground-truth values.

## 8. Summary and final remarks

We have described a methodology to estimate the redshift and physical properties of galaxies using broad-band photometry, in the context of *Euclid* preparation. The pipeline is designed to be agnostic with respect to the nature of the input catalogue and the properties to be estimated; a user may use the pipeline to estimate a variety of other properties for galaxies, or the properties of other classes of astronomical sources, provided a labeled tabular dataset is available.

**Table 3.** Metrics of model performance.

| | | COSMOS | Int Wide | | | | SED Wide | | | | EURISKO | | | | SPRITZ | | | |
|---|---|---|---|---|---|---|---|---|---|---|---|---|---|---|---|---|---|---|
| | | Case 0 | Case 1 | Case 2 | Case 3 | Case 4 | Case 1 | Case 2 | Case 3 | Case 4 | Case 1 | Case 2 | Case 3 | Case 4 | Case 1 | Case 2 | Case 3 | Case 4 |
| $z$ | NMAD | 0.029 | 0.103 | 0.09 | 0.024 | 0.02 | 0.108 | 0.082 | 0.031 | 0.023 | 0.022 | 0.025 | 0.004 | 0.004 | 0.139 | 0.112 | 0.085 | 0.052 |
| | $f_{\rm out}$ | 0.06 | 0.25 | 0.217 | 0.038 | 0.027 | 0.271 | 0.197 | 0.044 | 0.014 | 0.024 | 0.027 | 0.0004 | 0.0005 | 0.338 | 0.272 | 0.247 | 0.157 |
| | bias | -0.0005 | -0.0008 | 0.0006 | -0.0002 | -0.0001 | -0.0004 | -0.0004 | 0.0001 | -0.0001 | -0.0003 | -0.0004 | -0.0001 | -0.0001 | -0.0002 | 0.0011 | 0.0004 | 0.0002 |
| $M$ | NMAD | 0.119 | 0.178 | 0.167 | 0.088 | 0.078 | 0.222 | 0.192 | 0.134 | 0.116 | 0.115 | 0.112 | 0.03 | 0.031 | 0.351 | 0.288 | 0.255 | 0.189 |
| | $f_{\rm out}$ | 0.114 | 0.19 | 0.167 | 0.067 | 0.052 | 0.261 | 0.206 | 0.107 | 0.07 | 0.121 | 0.117 | 0.009 | 0.009 | 0.419 | 0.349 | 0.323 | 0.225 |
| | bias | -0.0016 | -0.003 | -0.001 | -0.003 | -0.004 | -0.0007 | -0.005 | -0.005 | -0.003 | -0.0004 | -0.0006 | -0.0003 | -0.0005 | $-7 \times 10^{-5}$ | 0.001 | -0.0004 | -0.0005 |
| | MAE | 0.161 | 0.212 | 0.195 | 0.121 | 0.108 | 0.266 | 0.231 | 0.156 | 0.126 | 0.142 | 0.144 | 0.036 | 0.037 | 0.372 | 0.319 | 0.308 | 0.237 |
| | $R^2$ | 0.83 | 0.759 | 0.776 | 0.87 | 0.877 | 0.677 | 0.717 | 0.866 | 0.911 | 0.907 | 0.895 | 0.992 | 0.991 | 0.642 | 0.703 | 0.71 | 0.791 |
| SFR | NMAD | 0.309 | 0.601 | 0.652 | 0.287 | 0.287 | 0.663 | 0.634 | 0.373 | 0.364 | 0.237 | 0.251 | 0.043 | 0.048 | 0.407 | 0.347 | 0.312 | 0.241 |
| | $f_{\rm out}$ | 0.379 | 0.605 | 0.625 | 0.354 | 0.347 | 0.632 | 0.618 | 0.435 | 0.427 | 0.311 | 0.317 | 0.035 | 0.04 | 0.47 | 0.414 | 0.383 | 0.303 |
| | bias | -0.0091 | -0.01 | -0.02 | -0.01 | -0.02 | -0.01 | -0.02 | -0.02 | -0.02 | -0.003 | -0.002 | -0.0008 | -0.002 | -0.0004 | 0.004 | -0.001 | 0.0005 |
| | MAE | 0.533 | 0.895 | 0.995 | 0.551 | 0.628 | 0.899 | 0.974 | 0.644 | 0.717 | 0.299 | 0.303 | 0.068 | 0.074 | 0.431 | 0.388 | 0.362 | 0.297 |
| | $R^2$ | 0.392 | 0.151 | 0.14 | 0.401 | 0.395 | 0.214 | 0.225 | 0.343 | 0.343 | 0.611 | 0.62 | 0.955 | 0.95 | 0.587 | 0.629 | 0.671 | 0.744 |
| sSFR | NMAD | 0.351 | 0.544 | 0.571 | 0.312 | 0.303 | 0.555 | 0.56 | 0.414 | 0.414 | 0.22 | 0.228 | 0.031 | 0.035 | 0.198 | 0.202 | 0.194 | 0.196 |
| | $f_{\rm out}$ | 0.421 | 0.57 | 0.586 | 0.388 | 0.378 | 0.576 | 0.579 | 0.476 | 0.475 | 0.319 | 0.317 | 0.042 | 0.044 | 0.141 | 0.155 | 0.127 | 0.133 |
| | bias | -0.009 | -0.004 | -0.001 | -0.01 | -0.02 | -0.006 | -0.01 | -0.02 | -0.02 | -0.0016 | 0.0002 | -0.0001 | -0.0003 | 0.001 | 0.002 | 0.0005 | 0.002 |
| | MAE | 0.526 | 0.808 | 0.912 | 0.541 | 0.615 | 0.779 | 0.886 | 0.655 | 0.733 | 0.296 | 0.287 | 0.064 | 0.067 | 0.187 | 0.193 | 0.176 | 0.177 |
| | $R^2$ | 0.423 | 0.227 | 0.214 | 0.431 | 0.435 | 0.303 | 0.299 | 0.368 | 0.384 | 0.692 | 0.677 | 0.965 | 0.958 | 0.156 | 0.184 | 0.26 | 0.325 |
| $E(B-V)$ | NMAD | 0.05 | 0.104 | 0.102 | 0.033 | 0.033 | 0.105 | 0.095 | 0.071 | 0.067 | 0.027 | 0.035 | 0.003 | 0.003 | | | | |
| | $f_{\rm out}$ | 0.02 | 0.022 | 0.025 | 0.01 | 0.008 | 0.027 | 0.026 | 0.02 | 0.018 | 0.012 | 0.007 | 0.0002 | 0.0003 | | | | |
| | bias | 2e-05 | 0.0006 | 0.0008 | 0.0001 | 0.0001 | 0.0002 | 0.0008 | 0.0009 | 0.001 | 0.0006 | 0.0006 | 0 | $-1 \times 10^{-5}$ | | | | |
| | MAE | 0.062 | 0.081 | 0.083 | 0.052 | 0.049 | 0.082 | 0.081 | 0.068 | 0.067 | 0.049 | 0.05 | 0.01 | 0.011 | | | | |
| | $R^2$ | 0.644 | 0.479 | 0.475 | 0.724 | 0.767 | 0.451 | 0.494 | 0.578 | 0.604 | 0.688 | 0.685 | 0.967 | 0.961 | | | | |
| Age | NMAD | 0.259 | 0.334 | 0.318 | 0.222 | 0.196 | 0.341 | 0.312 | 0.278 | 0.246 | 0.189 | 0.2 | 0.032 | 0.037 | | | | |
| | $f_{\rm out}$ | 0.297 | 0.391 | 0.37 | 0.253 | 0.222 | 0.399 | 0.358 | 0.308 | 0.27 | 0.288 | 0.286 | 0.038 | 0.038 | | | | |
| | bias | 0.001 | 0.002 | 0.004 | -0.002 | -0.004 | 0.006 | 0.002 | -0.001 | -0.003 | -0.0015 | -0.0011 | -0.0019 | -0.002 | | | | |
| | MAE | 0.239 | 0.299 | 0.287 | 0.209 | 0.191 | 0.307 | 0.29 | 0.246 | 0.225 | 0.24 | 0.234 | 0.061 | 0.064 | | | | |
| | $R^2$ | 0.565 | 0.342 | 0.35 | 0.653 | 0.686 | 0.306 | 0.342 | 0.541 | 0.58 | 0.681 | 0.651 | 0.966 | 0.957 | | | | |

**Table 4.** Example of the improvement in NMAD metric when using our pipeline, compared to a single regressor model, for Case 0.

| Label | Redshift | $M$ | SFR | sSFR | $E(B-V)$ | Age |
|---|---|---|---|---|---|---|
| Single | 0.0337 | 0.127 | 0.333 | 0.373 | 0.067 | 0.279 |
| Chained | 0.0291 | 0.118 | 0.313 | 0.352 | 0.048 | 0.260 |

The main novelty of our pipeline is its use of the `CatBoost` implementation of gradient-boosted regression-trees, together with chained regression and an intelligent, automatic optimization of the training data. We have shown that our chained regression is able to provide significantly better predictions for redshift and various physical properties, compared to when a single regressor is applied in isolation. In addition, we present a computationally efficient method to estimate the prediction uncertainties, and to predict performance metric values in the case where ground truth is not available.

In this paper, we have applied the pipeline to the problem of estimating the redshift and the following galaxy physical properies: log stellar mass ($M$), log star-formation rate (SFR), log specific star-formation rate (sSFR), $E(B-V)$, and log age. With the objective of evaluating the expected performance of our methodology for estimating the redshift and physical properties of galaxies imaged during the Euclid Wide Survey, we have applied our pipeline to several datasets consisting of mock *Euclid* broad-band photometry and mock LSST or UNIONS *ugriz* photometry: namely, Int Wide, SED Wide, EURISKO, and SPRITZ. We have evaluated the performance of our pipeline using normalized median absolute deviation (NMAD), catastrophic outlier fraction ($f_{\text{out}}$), and bias for redshift, or using NMAD, $f_{\text{out}}$, mean absolute error (MAE), and the $R^2$ score for physical properties.

We find that the metrics of model quality show a substantial dispersion across the range of mock *Euclid* catalogues used, and there is a strong dependence on whether only *Euclid* photometry, or *Euclid* and ancillary photometry, is used. In particular, the inclusion of ground-based optical photometry usually yields a very substantial improvement in the quality of the redshift and physical property estimates, despite some of these ancillary data containing non-detections. We also find that the construction methodology of the mock catalogues has a significant impact on the metric scores. In the interest of open science and reproducibility, we have also tested our pipeline using a subset of a publicly-available dataset, which we make available online.

For the application of our methodology to real photometry from *Euclid* and other large surveys, we envisage one of two main scenarios for the creation of a relevant training dataset. In the ideal case, one would select an area (or several areas) of the survey area for which high-quality, multiwavelength photometry and high-quality redshifts and physical properties estimates already exist. The training dataset would then be constructed by matching the existing redshift and physical property labels to the *Euclid* photometry. In the optical case, the training data would have the same noise properties as the test dataset for which the redshift and physical properties are to be predicted; in the event that the training data has significantly higher signal-to-noise, artificial scatter may be introduced to its photometry to mimic the lower quality of the test dataset.

A less ideal scenario, in the absence of suitable *Euclid* photometry, would be to follow a dataset creation methodology similar to that employed by Bisigello et al. (2020): photometry from a suitable area of sky is transformed to obtain expected broad-band magnitudes through the *Euclid* filters. In both cases, the complexity of real galaxy populations is preserved to a greater extent than in datasets constructed from template spectral energy distributions only.

Due to the sparsity of examples at $z \gtrsim 3.5$, the learning algorithm was unable to learn to reliably map the photometric information to the labels, rendering unreliable the predictions for redshift and the physical properties above this redshift. A potential solution for this issue would be to enlarge the training dataset such that the $z \gtrsim 3.5$ range is well populated. Additionally, using a more complex treatment of missing values, with missing photometry flagged differently depending on the cause (e.g., a non-detection versus no coverage), could plausibly be helpful, since it might allow information on the dropout of bluer bands at high-$z$ to be utilised more efficiently. Alternatively, traditional SED fitting could be used in this redshift regime.

Although we have tested our methodology using mock catalogues containing only galaxies without an AGN, we emphasise that there should not be any obstacle to the application of the methodology to other types of astrophysical objects or datasets. Provided suitable training data is available, our methodology could be applied to galaxies hosting an AGN, to stars, etc.

This paper is part of a wider project to develop and test methodologies for the estimation of galaxy redshift and physical properties using *Euclid* and ground-based photometry, as part of a 'data challenge' within the Euclid Collaboration (see also Euclid Collaboration: Bisigello et al. 2023). The scope of this paper is limited to presenting our new methodology and reporting its performance on several mock *Euclid* galaxy catalogues. A comparison between different physical property estimation methods are presented in a separate paper (Euclid Collaboration et al. 2024).

## Acknowledgments

We thank the anonymous A&A referee for feedback that helped to improve our manuscript. We also thank Karina Caputi for her thorough and helpful review of this manuscript as part of the internal Euclid Collaboration refereeing process. This work was supported by Fundação para a Ciência e a Tecnologia (FCT) through grants UID/FIS/04434/2019, UIDB/04434/2020, UIDP/04434/2020, and PTDC/FIS-AST/29245/2017, and an FCT-CAPES Transnational Coöperation Project. AH acknowledges support from the NVIDIA Academic Hardware Grant Program. PACC acknowledges financial support from the FCT through grant 2022.11477.BD.

The Euclid Consortium acknowledges the European Space Agency and a number of agencies and institutes that have supported the development of *Euclid*, in particular the Agenzia Spaziale Italiana, the Austrian Forschungsförderungsgesellschaft funded through BMK, the Belgian Science Policy, the Canadian Euclid Consortium, the Deutsches Zentrum für Luft- und Raumfahrt, the DTU Space and the Niels Bohr Institute in Denmark, the French Centre National d'Etudes Spatiales, the Fundação para a Ciência e a Tecnologia, the Hungarian Academy of Sciences, the Ministerio de Ciencia, Innovación y Universidades, the National Aeronautics and Space Administration, the National Astronomical Observatory of Japan, the Netherlandse Onderzoekschool Voor Astronomie, the Norwegian Space Agency, the Research Council of Finland, the Romanian Space Agency, the State Secretariat for Education, Research, and Innovation (SERI) at the Swiss Space Office (SSO), and the United Kingdom Space Agency. A complete and detailed list is available on the *Euclid* web site (www.euclid-ec.org).

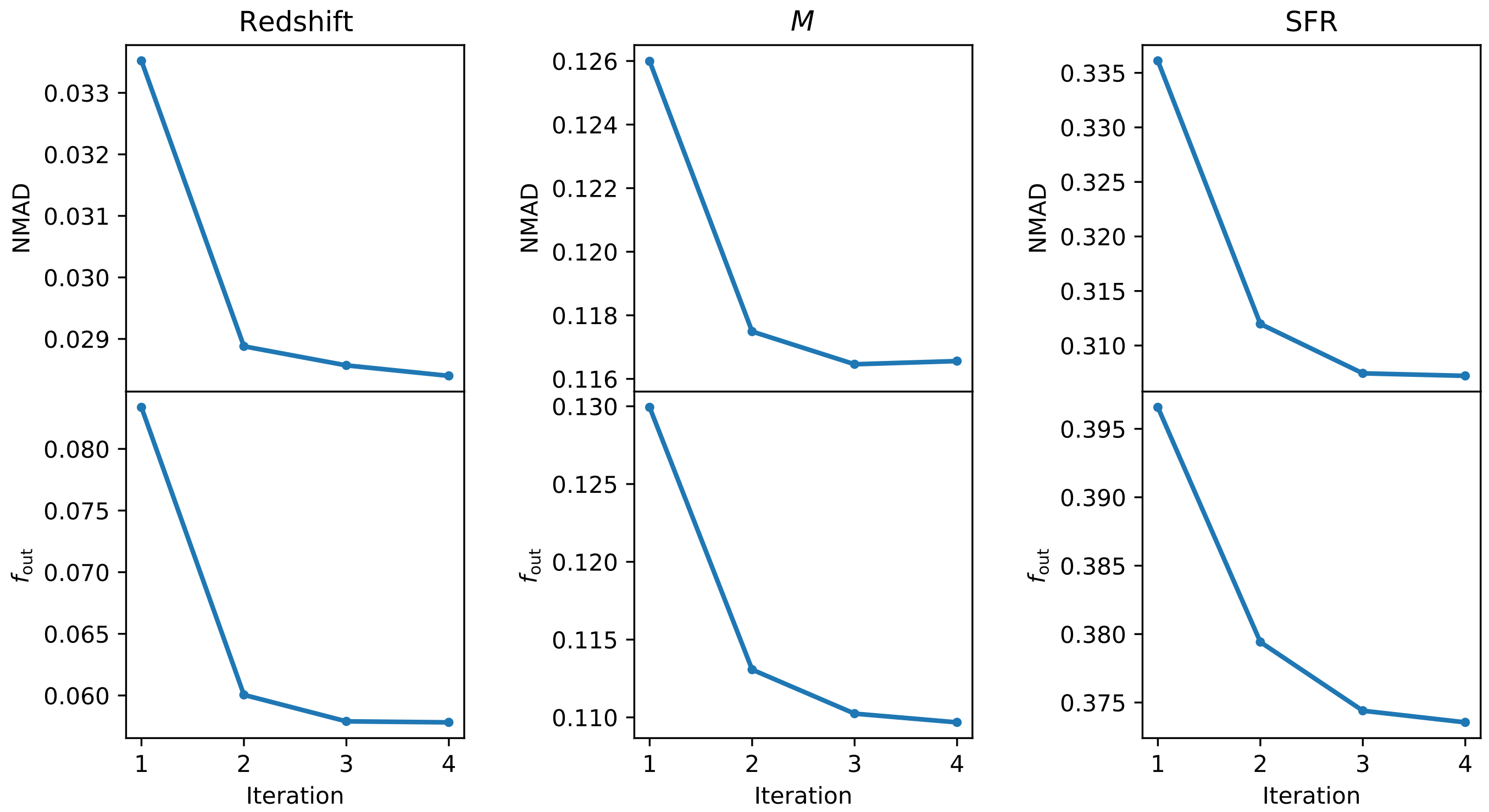

**Fig. 3.** The improvements in NMAD and $f_{\rm out}$ obtained after 4 iterations of our pipeline, when predicting redshift, $M$, and SFR for the COSMOS Case 0 dataset. For each of the physical properties, models with an awareness of the predicted values of the other properties make more accurate predictions, compared to models without it.

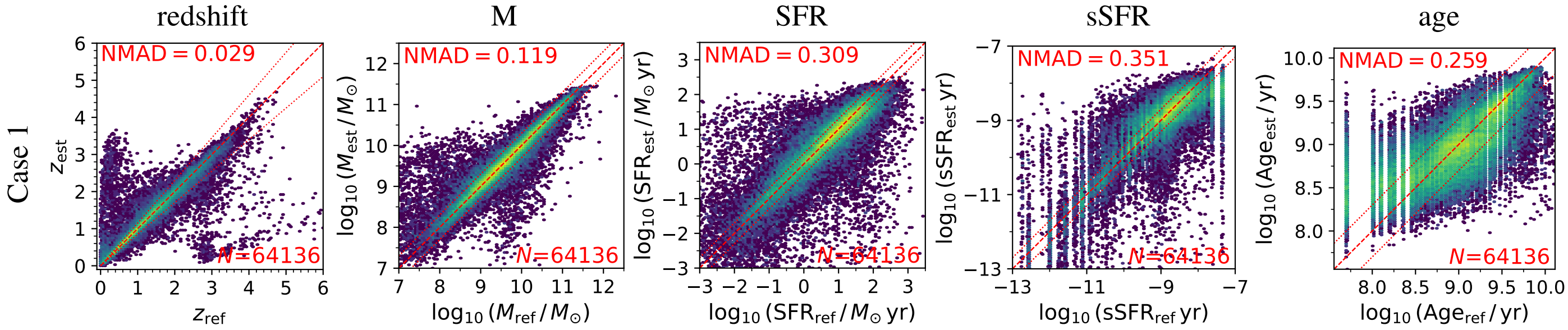

**Fig. 4.** Density maps showing estimated values versus the reference values for Redshift, $M$, SFR, sSFR and age for the COSMOS 2015 (Case 0) dataset. The dashed red line marks the case where the estimated value is equal to the reference value. The dotted red lines mark the area beyond which an estimated value is an outlier, using the criteria in Sect. 5. The vertical stripes visible in the sSFR and age results are caused by quantization of these properties in the ground-truth labels.

Based on data products from observations made with ESO Telescopes at the La Silla Paranal Observatory under ESO programme ID 179.A-2005 and on data products produced by TERAPIX and the Cambridge Astronomy Survey Unit on behalf of the UltraVISTA consortium.

In the development of our pipeline, we have made use of the `Scikit-Learn` (Pedregosa et al. 2011), `Pandas` (McKinney 2010), `Numpy` (Harris et al. 2020), `Scipy` (Virtanen et al. 2020), `Dask` (Rocklin 2015), and `CatBoost` (Prokhorenkova et al. 2018) packages for `cPython`.

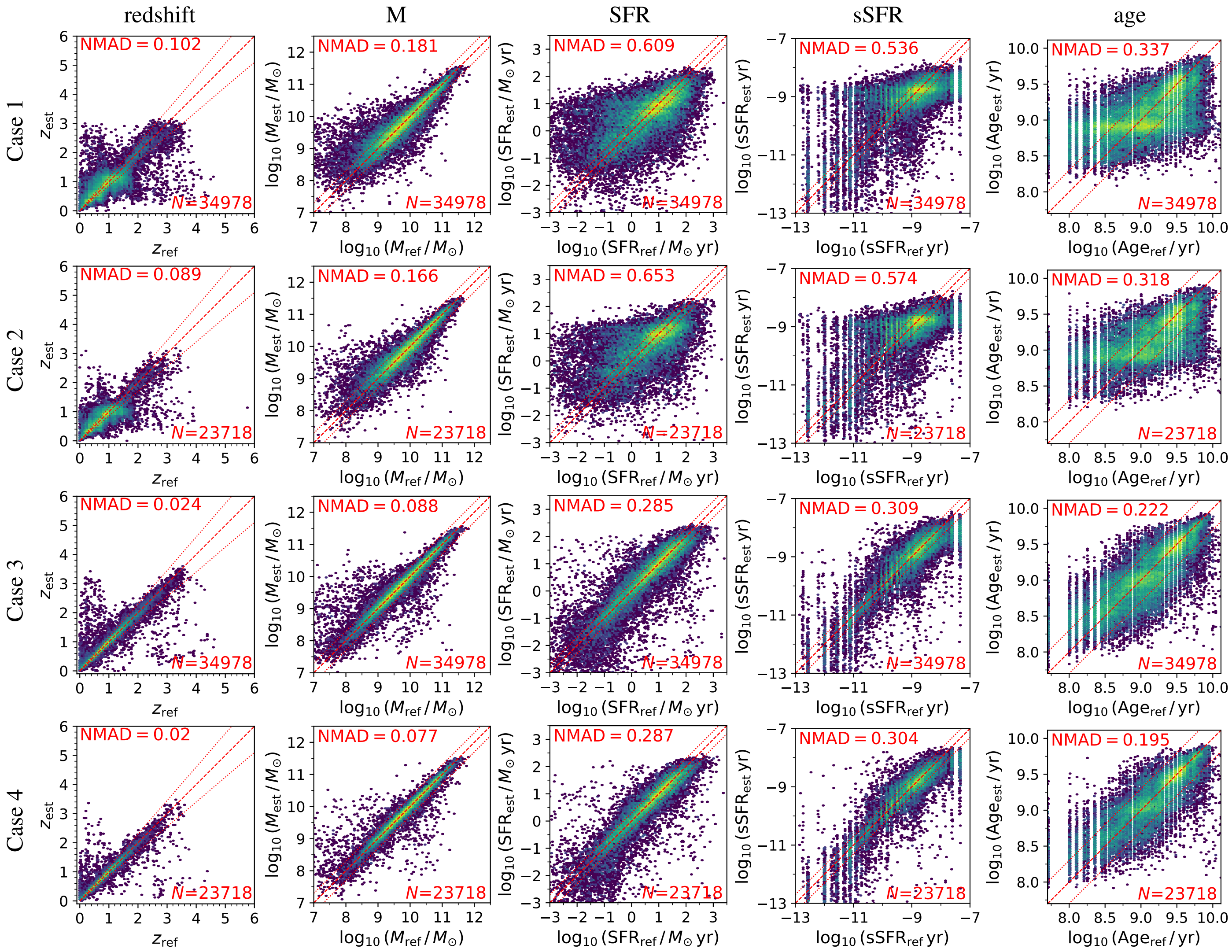

**Fig. 5.** Density maps showing estimated values versus the reference values for Redshift, $M$, SFR, sSFR and age for the Int Wide mock *Euclid* catalogue. Showing are Case 1 (first row), Case 2 (second row), Case 3 (third row), and Case 4 (fourth row). The dashed red line marks the case where the estimated value is equal to the reference value. The dotted red lines mark the area beyond which an estimated value is an outlier, using the criteria in Sect. 5. The vertical stripes visible in the sSFR and age results are caused by quantization of these properties in the ground-truth labels.

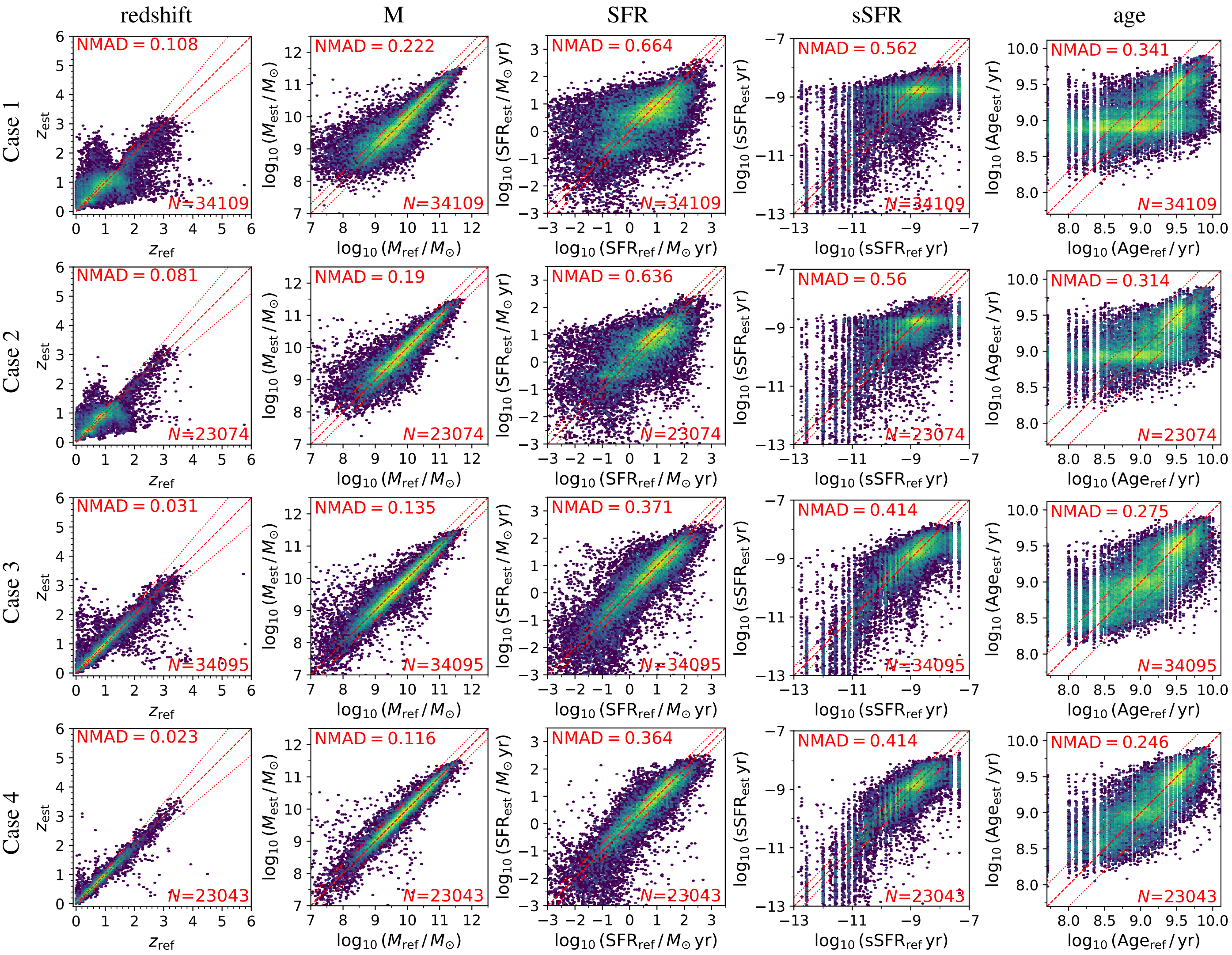

**Fig. 6.** Density maps showing estimated values versus the reference values for Redshift, $M$, SFR, sSFR and age for the SED Wide mock *Euclid* catalogue. Showing are Case 1 (first row), Case 2 (second row), Case 3 (third row), and Case 4 (fourth row). The dashed red line marks the case where the estimated value is equal to the reference value. The dotted red lines mark the area beyond which an estimated value is an outlier, using the criteria in Sect. 5. The vertical stripes visible in the sSFR and age results are caused by quantization of these properties in the ground-truth labels.

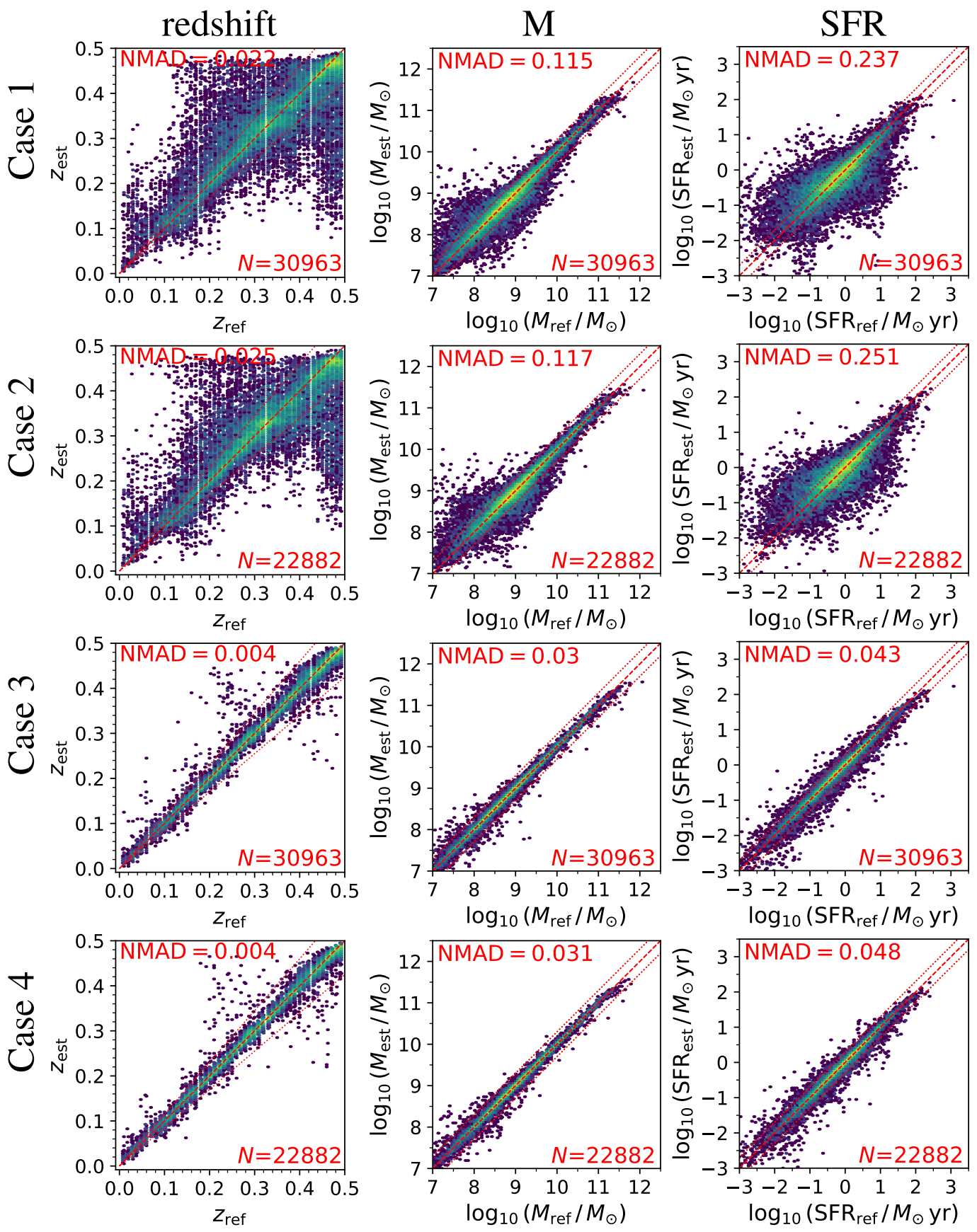

**Fig. 7.** Density maps showing estimated values versus the reference values for Redshift, $M$, and SFR for the EURISKO mock *Euclid* catalogue. Showing are Case 1 (first row), Case 2 (second row), Case 3 (third row), and Case 4 (fourth row). The dashed red line marks the case where the estimated value is equal to the reference value. The dotted red lines mark the area beyond which an estimated value is an outlier, using the criteria in Sect. 5.

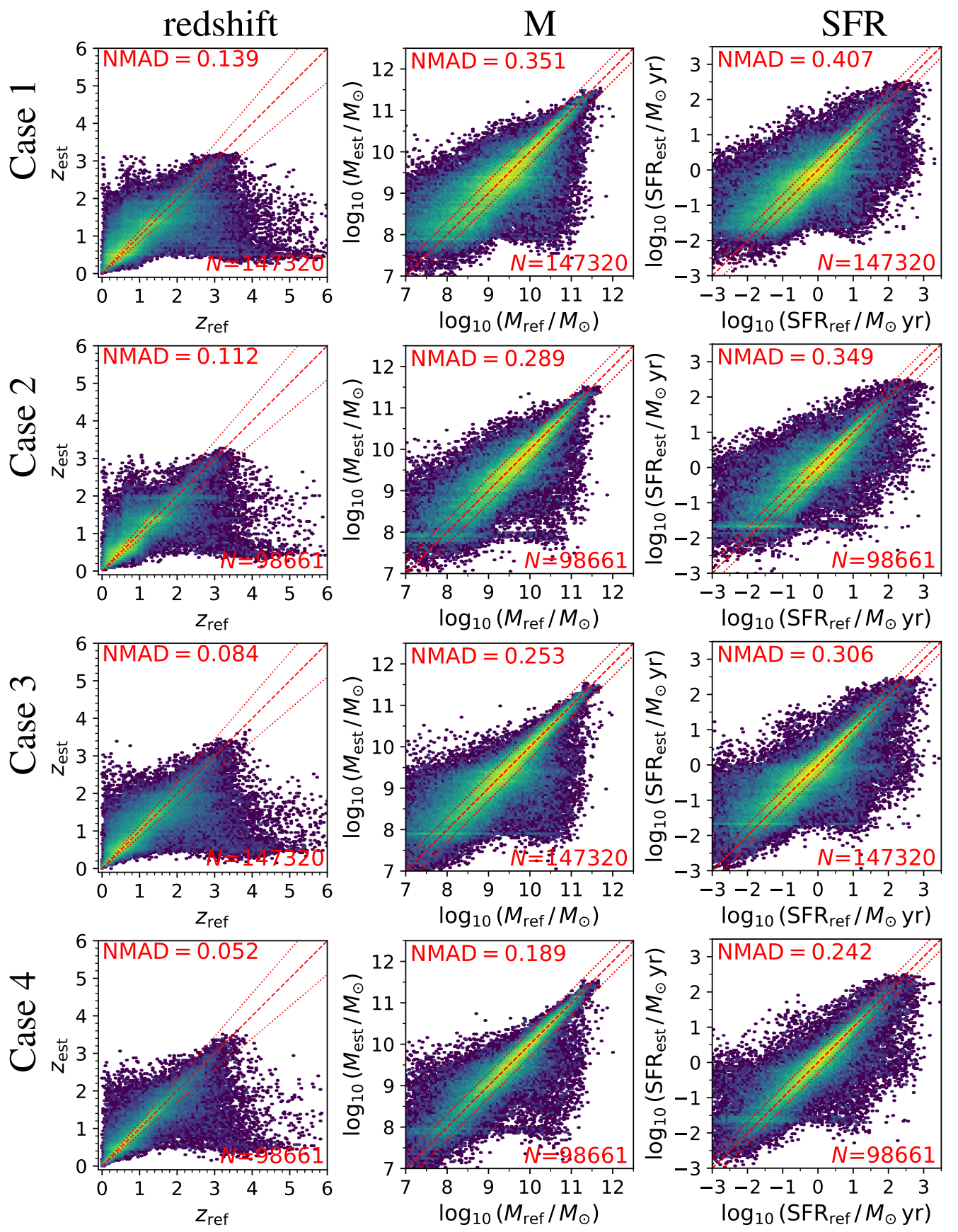

**Fig. 8.** Density maps showing estimated values versus the reference values for Redshift, $M$, and SFR for the SPRITZ mock *Euclid* catalogue. Showing are Case 1 (first row), Case 2 (second row), Case 3 (third row), and Case 4 (fourth row). The dashed red line marks the case where the estimated value is equal to the reference value. The dotted red lines mark the area beyond which an estimated value is an outlier, using the criteria in Sect. 5.

[1] Instituto de Astrofísica e Ciências do Espaço, Universidade do Porto, CAUP, Rua das Estrelas, PT4150-762 Porto, Portugal
[2] DTx – Digital Transformation CoLAB, Building 1, Azurém Campus, University of Minho, 4800-058 Guimarães, Portugal
[3] Faculdade de Ciências da Universidade do Porto, Rua do Campo de Alegre, 4150-007 Porto, Portugal
[4] INAF, Istituto di Radioastronomia, Via Piero Gobetti 101, 40129 Bologna, Italy
[5] Dipartimento di Fisica e Astronomia "G. Galilei", Università di Padova, Via Marzolo 8, 35131 Padova, Italy
[6] INAF-Osservatorio Astronomico di Capodimonte, Via Moiariello 16, 80131 Napoli, Italy
[7] INAF-Osservatorio di Astrofisica e Scienza dello Spazio di Bologna, Via Piero Gobetti 93/3, 40129 Bologna, Italy
[8] Sterrenkundig Observatorium, Universiteit Gent, Krijgslaan 281 S9, 9000 Gent, Belgium
[9] INAF-Osservatorio Astronomico di Brera, Via Brera 28, 20122 Milano, Italy
[10] School of Mathematics and Physics, University of Surrey, Guildford, Surrey, GU2 7XH, UK
[11] SISSA, International School for Advanced Studies, Via Bonomea 265, 34136 Trieste TS, Italy
[12] INAF-Osservatorio Astronomico di Trieste, Via G. B. Tiepolo 11, 34143 Trieste, Italy
[13] INFN, Sezione di Trieste, Via Valerio 2, 34127 Trieste TS, Italy
[14] IFPU, Institute for Fundamental Physics of the Universe, via Beirut 2, 34151 Trieste, Italy
[15] Dipartimento di Fisica e Astronomia, Università di Bologna, Via Gobetti 93/2, 40129 Bologna, Italy
[16] INFN-Sezione di Bologna, Viale Berti Pichat 6/2, 40127 Bologna, Italy
[17] Max Planck Institute for Extraterrestrial Physics, Giessenbachstr. 1, 85748 Garching, Germany
[18] INAF-Osservatorio Astrofisico di Torino, Via Osservatorio 20, 10025 Pino Torinese (TO), Italy
[19] Dipartimento di Fisica, Università di Genova, Via Dodecaneso 33, 16146, Genova, Italy
[20] INFN-Sezione di Genova, Via Dodecaneso 33, 16146, Genova, Italy
[21] Department of Physics "E. Pancini", University Federico II, Via Cinthia 6, 80126, Napoli, Italy
[22] INFN section of Naples, Via Cinthia 6, 80126, Napoli, Italy
[23] Dipartimento di Fisica, Università degli Studi di Torino, Via P. Giuria 1, 10125 Torino, Italy

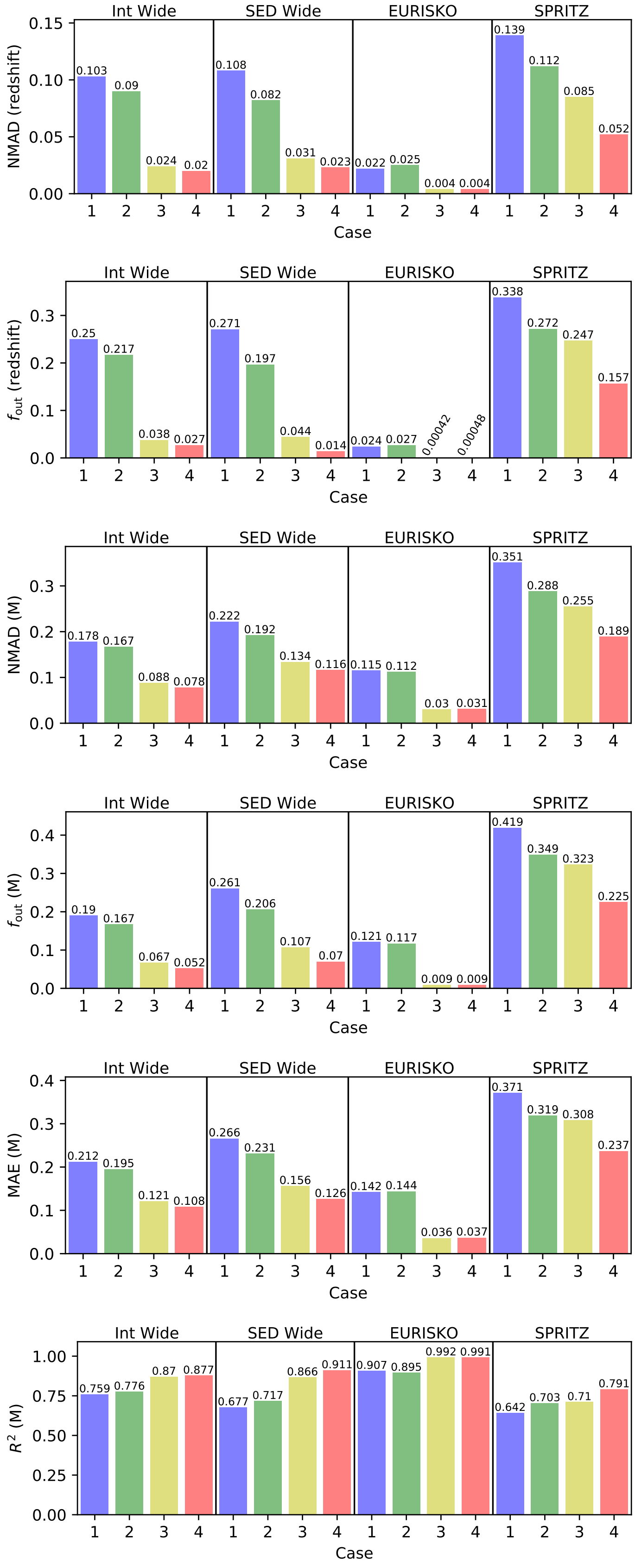

**Fig. 9.** Bar charts showing the NMAD, $f_{\rm out}$, MAE, and $R^2$ metrics for the $z$ and $M$ predictions. The $x$-axis separates the results by Case and catalogue.

[24] INFN-Sezione di Torino, Via P. Giuria 1, 10125 Torino, Italy
[25] INAF-IASF Milano, Via Alfonso Corti 12, 20133 Milano, Italy
[26] Institut de Física d'Altes Energies (IFAE), The Barcelona Institute of Science and Technology, Campus UAB, 08193 Bellaterra (Barcelona), Spain
[27] Port d'Informació Científica, Campus UAB, C. Albareda s/n, 08193 Bellaterra (Barcelona), Spain
[28] Institute for Theoretical Particle Physics and Cosmology (TTK), RWTH Aachen University, 52056 Aachen, Germany
[29] INAF-Osservatorio Astronomico di Roma, Via Frascati 33, 00078 Monteporzio Catone, Italy
[30] Dipartimento di Fisica e Astronomia "Augusto Righi" - Alma Mater Studiorum Università di Bologna, via Piero Gobetti 93/2, 40129 Bologna, Italy
[31] Dipartimento di Fisica e Astronomia "Augusto Righi" - Alma Mater Studiorum Università di Bologna, Viale Berti Pichat 6/2, 40127 Bologna, Italy
[32] Instituto de Astrofísica de Canarias, Calle Vía Láctea s/n, 38204, San Cristóbal de La Laguna, Tenerife, Spain
[33] Institute for Astronomy, University of Edinburgh, Royal Observatory, Blackford Hill, Edinburgh EH9 3HJ, UK
[34] Jodrell Bank Centre for Astrophysics, Department of Physics and Astronomy, University of Manchester, Oxford Road, Manchester M13 9PL, UK
[35] European Space Agency/ESRIN, Largo Galileo Galilei 1, 00044 Frascati, Roma, Italy
[36] ESAC/ESA, Camino Bajo del Castillo, s/n., Urb. Villafranca del Castillo, 28692 Villanueva de la Cañada, Madrid, Spain
[37] Université Claude Bernard Lyon 1, CNRS/IN2P3, IP2I Lyon, UMR 5822, Villeurbanne, F-69100, France
[38] Institute of Physics, Laboratory of Astrophysics, Ecole Polytechnique Fédérale de Lausanne (EPFL), Observatoire de Sauverny, 1290 Versoix, Switzerland
[39] UCB Lyon 1, CNRS/IN2P3, IUF, IP2I Lyon, 4 rue Enrico Fermi, 69622 Villeurbanne, France
[40] Departamento de Física, Faculdade de Ciências, Universidade de Lisboa, Edifício C8, Campo Grande, PT1749-016 Lisboa, Portugal
[41] Instituto de Astrofísica e Ciências do Espaço, Faculdade de Ciências, Universidade de Lisboa, Campo Grande, 1749-016 Lisboa, Portugal
[42] Department of Astronomy, University of Geneva, ch. d'Ecogia 16, 1290 Versoix, Switzerland
[43] INFN-Padova, Via Marzolo 8, 35131 Padova, Italy
[44] INAF-Istituto di Astrofisica e Planetologia Spaziali, via del Fosso del Cavaliere, 100, 00100 Roma, Italy
[45] Université Paris-Saclay, Université Paris Cité, CEA, CNRS, AIM, 91191, Gif-sur-Yvette, France
[46] Universitäts-Sternwarte München, Fakultät für Physik, Ludwig-Maximilians-Universität München, Scheinerstrasse 1, 81679 München, Germany
[47] Istituto Nazionale di Fisica Nucleare, Sezione di Bologna, Via Irnerio 46, 40126 Bologna, Italy
[48] INAF-Osservatorio Astronomico di Padova, Via dell'Osservatorio 5, 35122 Padova, Italy
[49] Dipartimento di Fisica "Aldo Pontremoli", Università degli Studi di Milano, Via Celoria 16, 20133 Milano, Italy
[50] INFN-Sezione di Milano, Via Celoria 16, 20133 Milano, Italy
[51] Institute of Theoretical Astrophysics, University of Oslo, P.O. Box 1029 Blindern, 0315 Oslo, Norway
[52] Jet Propulsion Laboratory, California Institute of Technology, 4800 Oak Grove Drive, Pasadena, CA, 91109, USA
[53] Department of Physics, Lancaster University, Lancaster, LA1 4YB, UK
[54] von Hoerner & Sulger GmbH, Schlossplatz 8, 68723 Schwetzingen, Germany
[55] Technical University of Denmark, Elektrovej 327, 2800 Kgs. Lyngby, Denmark
[56] Cosmic Dawn Center (DAWN), Denmark
[57] Max-Planck-Institut für Astronomie, Königstuhl 17, 69117 Heidelberg, Germany

[58] Department of Physics and Astronomy, University College London, Gower Street, London WC1E 6BT, UK
[59] Department of Physics and Helsinki Institute of Physics, Gustaf Hällströmin katu 2, 00014 University of Helsinki, Finland
[60] Aix-Marseille Université, CNRS/IN2P3, CPPM, Marseille, France
[61] Université de Genève, Département de Physique Théorique and Centre for Astroparticle Physics, 24 quai Ernest-Ansermet, CH-1211 Genève 4, Switzerland
[62] Department of Physics, P.O. Box 64, 00014 University of Helsinki, Finland
[63] Helsinki Institute of Physics, Gustaf Hällströmin katu 2, University of Helsinki, Helsinki, Finland
[64] NOVA optical infrared instrumentation group at ASTRON, Oude Hoogeveensedijk 4, 7991PD, Dwingeloo, The Netherlands
[65] Centre de Calcul de l'IN2P3/CNRS, 21 avenue Pierre de Coubertin 69627 Villeurbanne Cedex, France
[66] Universität Bonn, Argelander-Institut für Astronomie, Auf dem Hügel 71, 53121 Bonn, Germany
[67] INFN-Sezione di Roma, Piazzale Aldo Moro, 2 - c/o Dipartimento di Fisica, Edificio G. Marconi, 00185 Roma, Italy
[68] Aix-Marseille Université, CNRS, CNES, LAM, Marseille, France
[69] Department of Physics, Centre for Extragalactic Astronomy, Durham University, South Road, DH1 3LE, UK
[70] Institut d'Astrophysique de Paris, UMR 7095, CNRS, and Sorbonne Université, 98 bis boulevard Arago, 75014 Paris, France
[71] Université Paris Cité, CNRS, Astroparticule et Cosmologie, 75013 Paris, France
[72] University of Applied Sciences and Arts of Northwestern Switzerland, School of Engineering, 5210 Windisch, Switzerland
[73] Institut d'Astrophysique de Paris, 98bis Boulevard Arago, 75014, Paris, France
[74] European Space Agency/ESTEC, Keplerlaan 1, 2201 AZ Noordwijk, The Netherlands
[75] School of Mathematics, Statistics and Physics, Newcastle University, Herschel Building, Newcastle-upon-Tyne, NE1 7RU, UK
[76] Department of Physics, Institute for Computational Cosmology, Durham University, South Road, DH1 3LE, UK
[77] Department of Physics and Astronomy, University of Aarhus, Ny Munkegade 120, DK-8000 Aarhus C, Denmark
[78] Space Science Data Center, Italian Space Agency, via del Politecnico snc, 00133 Roma, Italy
[79] Centre National d'Etudes Spatiales – Centre spatial de Toulouse, 18 avenue Edouard Belin, 31401 Toulouse Cedex 9, France
[80] Institute of Space Science, Str. Atomistilor, nr. 409 Măgurele, Ilfov, 077125, Romania
[81] Departamento de Astrofísica, Universidad de La Laguna, 38206, La Laguna, Tenerife, Spain
[82] Institut für Theoretische Physik, University of Heidelberg, Philosophenweg 16, 69120 Heidelberg, Germany
[83] Institut de Recherche en Astrophysique et Planétologie (IRAP), Université de Toulouse, CNRS, UPS, CNES, 14 Av. Edouard Belin, 31400 Toulouse, France
[84] Université St Joseph; Faculty of Sciences, Beirut, Lebanon
[85] Departamento de Física, FCFM, Universidad de Chile, Blanco Encalada 2008, Santiago, Chile
[86] Universität Innsbruck, Institut für Astro- und Teilchenphysik, Technikerstr. 25/8, 6020 Innsbruck, Austria
[87] Institut d'Estudis Espacials de Catalunya (IEEC), Edifici RDIT, Campus UPC, 08860 Castelldefels, Barcelona, Spain
[88] Institute of Space Sciences (ICE, CSIC), Campus UAB, Carrer de Can Magrans, s/n, 08193 Barcelona, Spain
[89] Satlantis, University Science Park, Sede Bld 48940, Leioa-Bilbao, Spain
[90] Centro de Investigaciones Energéticas, Medioambientales y Tecnológicas (CIEMAT), Avenida Complutense 40, 28040 Madrid, Spain
[91] Instituto de Astrofísica e Ciências do Espaço, Faculdade de Ciências, Universidade de Lisboa, Tapada da Ajuda, 1349-018 Lisboa, Portugal
[92] Universidad Politécnica de Cartagena, Departamento de Electrónica y Tecnología de Computadoras, Plaza del Hospital 1, 30202 Cartagena, Spain
[93] INFN-Bologna, Via Irnerio 46, 40126 Bologna, Italy
[94] Infrared Processing and Analysis Center, California Institute of Technology, Pasadena, CA 91125, USA
[95] Astronomical Observatory of the Autonomous Region of the Aosta Valley (OAVdA), Loc. Lignan 39, I-11020, Nus (Aosta Valley), Italy
[96] Junia, EPA department, 41 Bd Vauban, 59800 Lille, France
[97] ICSC - Centro Nazionale di Ricerca in High Performance Computing, Big Data e Quantum Computing, Via Magnanelli 2, Bologna, Italy
[98] Instituto de Física Teórica UAM-CSIC, Campus de Cantoblanco, 28049 Madrid, Spain
[99] CERCA/ISO, Department of Physics, Case Western Reserve University, 10900 Euclid Avenue, Cleveland, OH 44106, USA
[100] Laboratoire Univers et Théorie, Observatoire de Paris, Université PSL, Université Paris Cité, CNRS, 92190 Meudon, France
[101] Dipartimento di Fisica e Scienze della Terra, Università degli Studi di Ferrara, Via Giuseppe Saragat 1, 44122 Ferrara, Italy
[102] Istituto Nazionale di Fisica Nucleare, Sezione di Ferrara, Via Giuseppe Saragat 1, 44122 Ferrara, Italy
[103] Dipartimento di Fisica - Sezione di Astronomia, Università di Trieste, Via Tiepolo 11, 34131 Trieste, Italy
[104] Minnesota Institute for Astrophysics, University of Minnesota, 116 Church St SE, Minneapolis, MN 55455, USA
[105] Institute Lorentz, Leiden University, Niels Bohrweg 2, 2333 CA Leiden, The Netherlands
[106] Université Côte d'Azur, Observatoire de la Côte d'Azur, CNRS, Laboratoire Lagrange, Bd de l'Observatoire, CS 34229, 06304 Nice cedex 4, France
[107] Institute for Astronomy, University of Hawaii, 2680 Woodlawn Drive, Honolulu, HI 96822, USA
[108] Department of Physics & Astronomy, University of California Irvine, Irvine CA 92697, USA
[109] Department of Astronomy & Physics and Institute for Computational Astrophysics, Saint Mary's University, 923 Robie Street, Halifax, Nova Scotia, B3H 3C3, Canada
[110] Departamento Física Aplicada, Universidad Politécnica de Cartagena, Campus Muralla del Mar, 30202 Cartagena, Murcia, Spain
[111] Department of Physics, Oxford University, Keble Road, Oxford OX1 3RH, UK
[112] Institute of Cosmology and Gravitation, University of Portsmouth, Portsmouth PO1 3FX, UK
[113] Department of Computer Science, Aalto University, PO Box 15400, Espoo, FI-00 076, Finland
[114] Ruhr University Bochum, Faculty of Physics and Astronomy, Astronomical Institute (AIRUB), German Centre for Cosmological Lensing (GCCL), 44780 Bochum, Germany
[115] DARK, Niels Bohr Institute, University of Copenhagen, Jagtvej 155, 2200 Copenhagen, Denmark
[116] Department of Physics and Astronomy, Vesilinnantie 5, 20014 University of Turku, Finland
[117] Serco for European Space Agency (ESA), Camino bajo del Castillo, s/n, Urbanizacion Villafranca del Castillo, Villanueva de la Cañada, 28692 Madrid, Spain
[118] ARC Centre of Excellence for Dark Matter Particle Physics, Melbourne, Australia
[119] Centre for Astrophysics & Supercomputing, Swinburne University of Technology, Hawthorn, Victoria 3122, Australia
[120] W.M. Keck Observatory, 65-1120 Mamalahoa Hwy, Kamuela, HI, USA
[121] School of Physics and Astronomy, Queen Mary University of London, Mile End Road, London E1 4NS, UK
[122] Department of Physics and Astronomy, University of the Western Cape, Bellville, Cape Town, 7535, South Africa
[123] ICTP South American Institute for Fundamental Research, Instituto de Física Teórica, Universidade Estadual Paulista, São Paulo, Brazil

[124] Oskar Klein Centre for Cosmoparticle Physics, Department of Physics, Stockholm University, Stockholm, SE-106 91, Sweden
[125] Astrophysics Group, Blackett Laboratory, Imperial College London, London SW7 2AZ, UK
[126] INAF-Osservatorio Astrofisico di Arcetri, Largo E. Fermi 5, 50125, Firenze, Italy
[127] Dipartimento di Fisica, Sapienza Università di Roma, Piazzale Aldo Moro 2, 00185 Roma, Italy
[128] Centro de Astrofísica da Universidade do Porto, Rua das Estrelas, 4150-762 Porto, Portugal
[129] Université Paris-Saclay, CNRS, Institut d'astrophysique spatiale, 91405, Orsay, France
[130] Institute of Astronomy, University of Cambridge, Madingley Road, Cambridge CB3 0HA, UK
[131] Univ. Grenoble Alpes, CNRS, Grenoble INP, LPSC-IN2P3, 53, Avenue des Martyrs, 38000, Grenoble, France
[132] Department of Astrophysics, University of Zurich, Winterthurerstrasse 190, 8057 Zurich, Switzerland
[133] Dipartimento di Fisica, Università degli studi di Genova, and INFN-Sezione di Genova, via Dodecaneso 33, 16146, Genova, Italy
[134] Theoretical astrophysics, Department of Physics and Astronomy, Uppsala University, Box 515, 751 20 Uppsala, Sweden
[135] Mullard Space Science Laboratory, University College London, Holmbury St Mary, Dorking, Surrey RH5 6NT, UK
[136] Department of Astrophysical Sciences, Peyton Hall, Princeton University, Princeton, NJ 08544, USA
[137] Niels Bohr Institute, University of Copenhagen, Jagtvej 128, 2200 Copenhagen, Denmark
[138] Cosmic Dawn Center (DAWN)
[139] Center for Cosmology and Particle Physics, Department of Physics, New York University, New York, NY 10003, USA
[140] Center for Computational Astrophysics, Flatiron Institute, 162 5th Avenue, 10010, New York, NY, USA

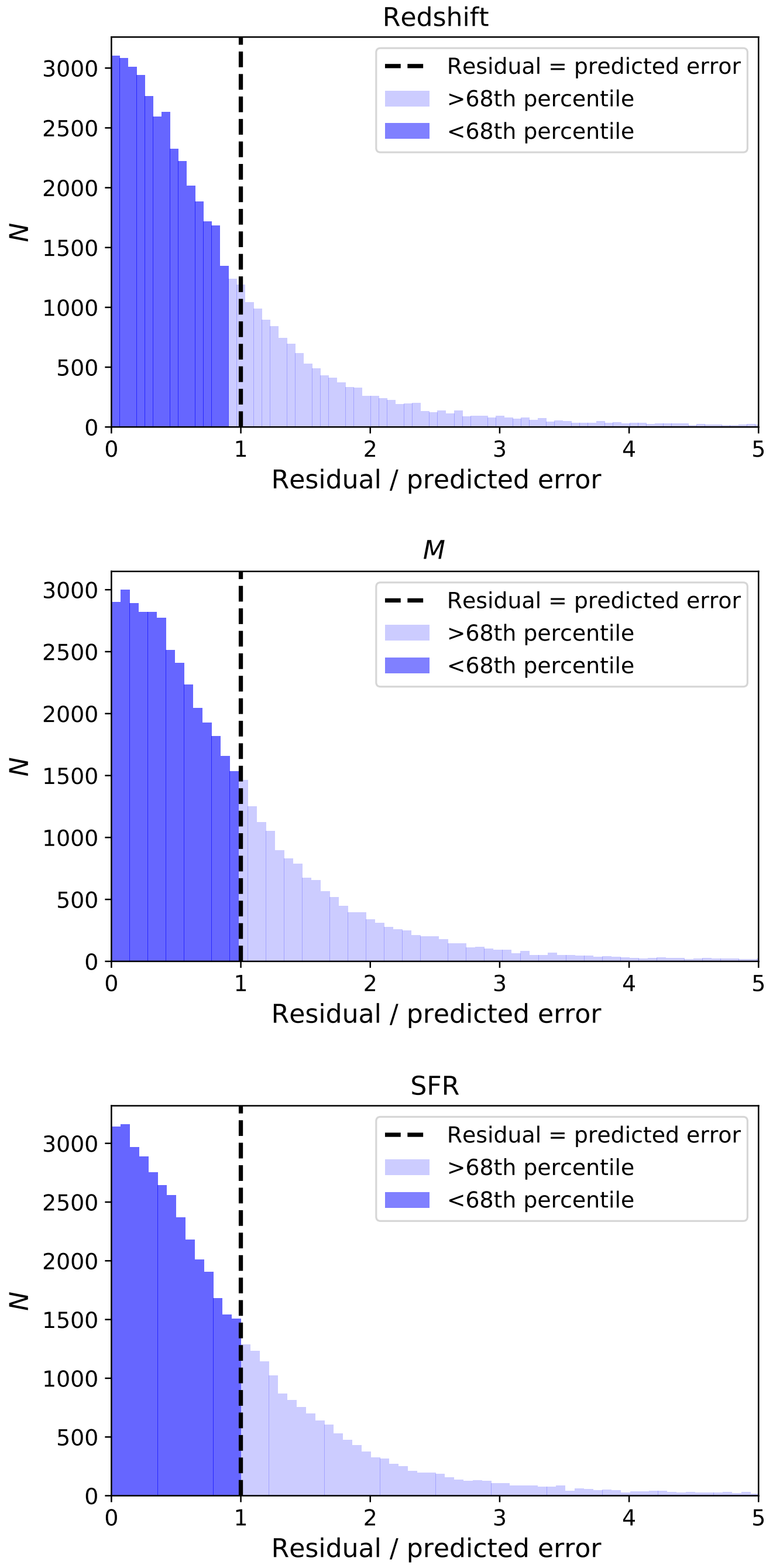

**Fig. A.1.** Histograms showing the distribution of residuals with respect to the predicted 68 % confidence interval, when predicting redshift, $M$ or SFR using the Int Wide catalogue with the Case 4 configuration.

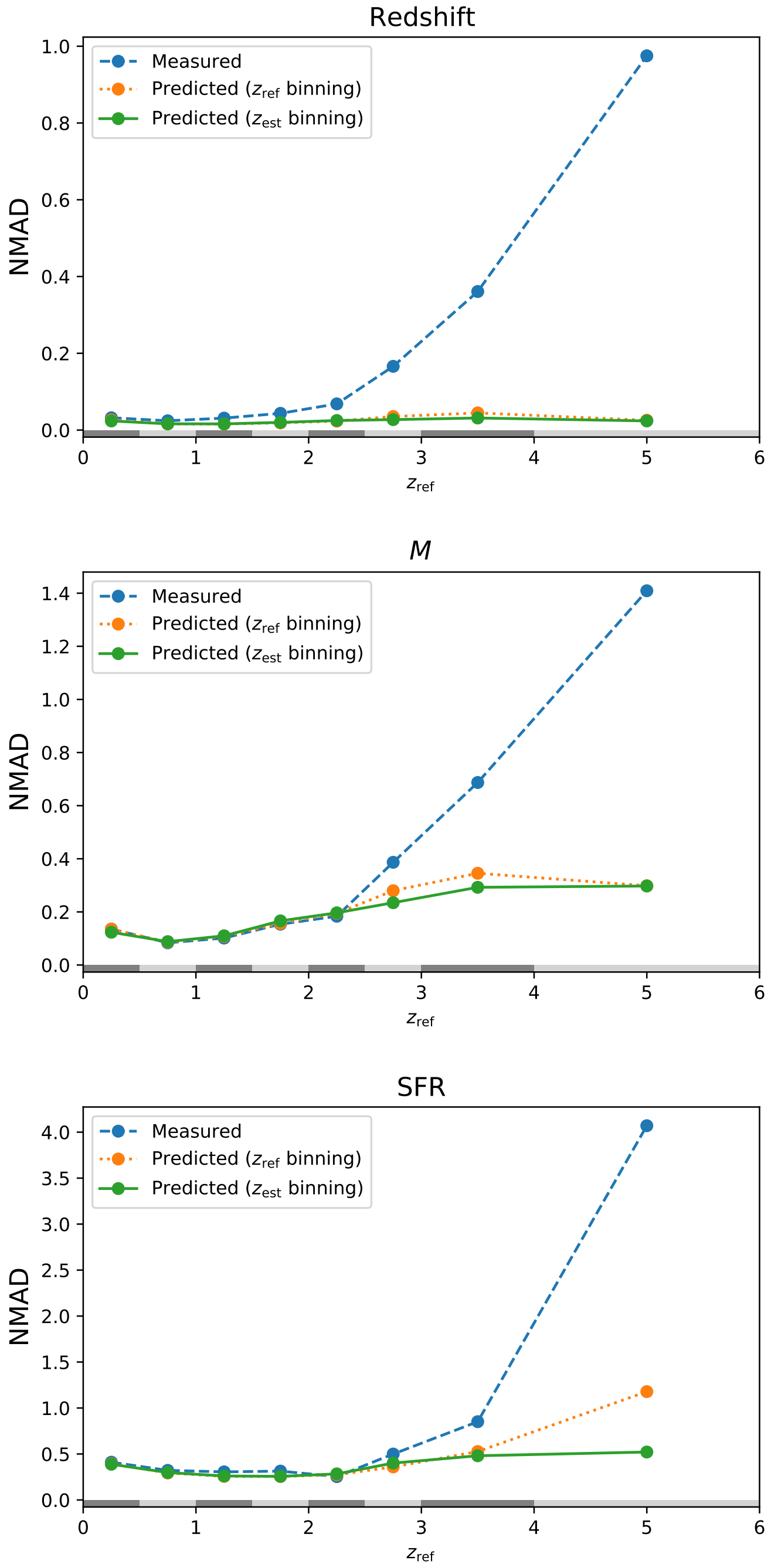

**Fig. A.2.** Testing the performance of our error estimation method in different redshift bins, for the Int Wide catalogue (Case 4). The dashed blue line shows the true NMAD values; the lines shows the NMAD values calculated using our error estimates, with redshift binning performed using the ground-truth ($z_{ref}$; orange dotted line), with the redshift binning done using the estimated redshifts ($z_{est}$; solid green line). The grey rectangles just above the $x$-axis indicate the range of redshift covered by the bins.

## Appendix A: Supplementary figures

Here, we show a number of supplementary figures that are referred to in the main text of this paper.

### *Appendix A.1: Uncertainty and performance estimation*

In this appendix, we show supplementary figures related to the estimation of prediction uncertainties (Fig. A.1), and the estimation of model performance in the absence of ground truth labels (Fig. A.2), referred to in Sect. 6.5.

### *Appendix A.2: Additional figures*

In this Appendix, we present supplementary figures referred to in Sect. 7.3.

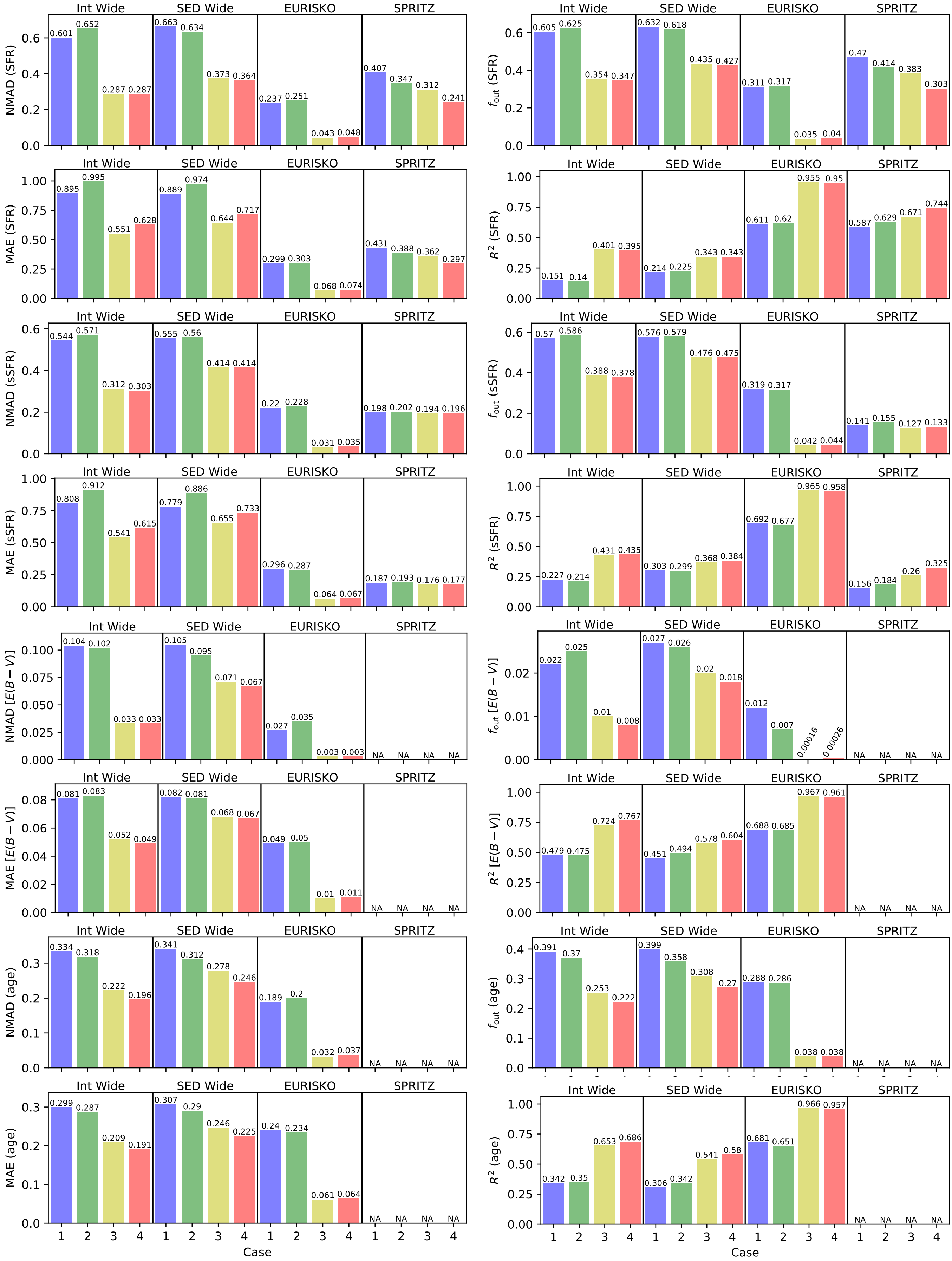

**Fig. A.3.** Similar to Fig. 9: Bar charts showing the NMAD, $f_{\rm out}$, MAE, and $R^2$ metrics for the predictions of SFR, sSFR, $E(B-V)$, and age. The $x$-axis separates the results by Case and catalogue. 'NA' indicates that a quantity was not among the predicted labels for that particular mock catalogue.

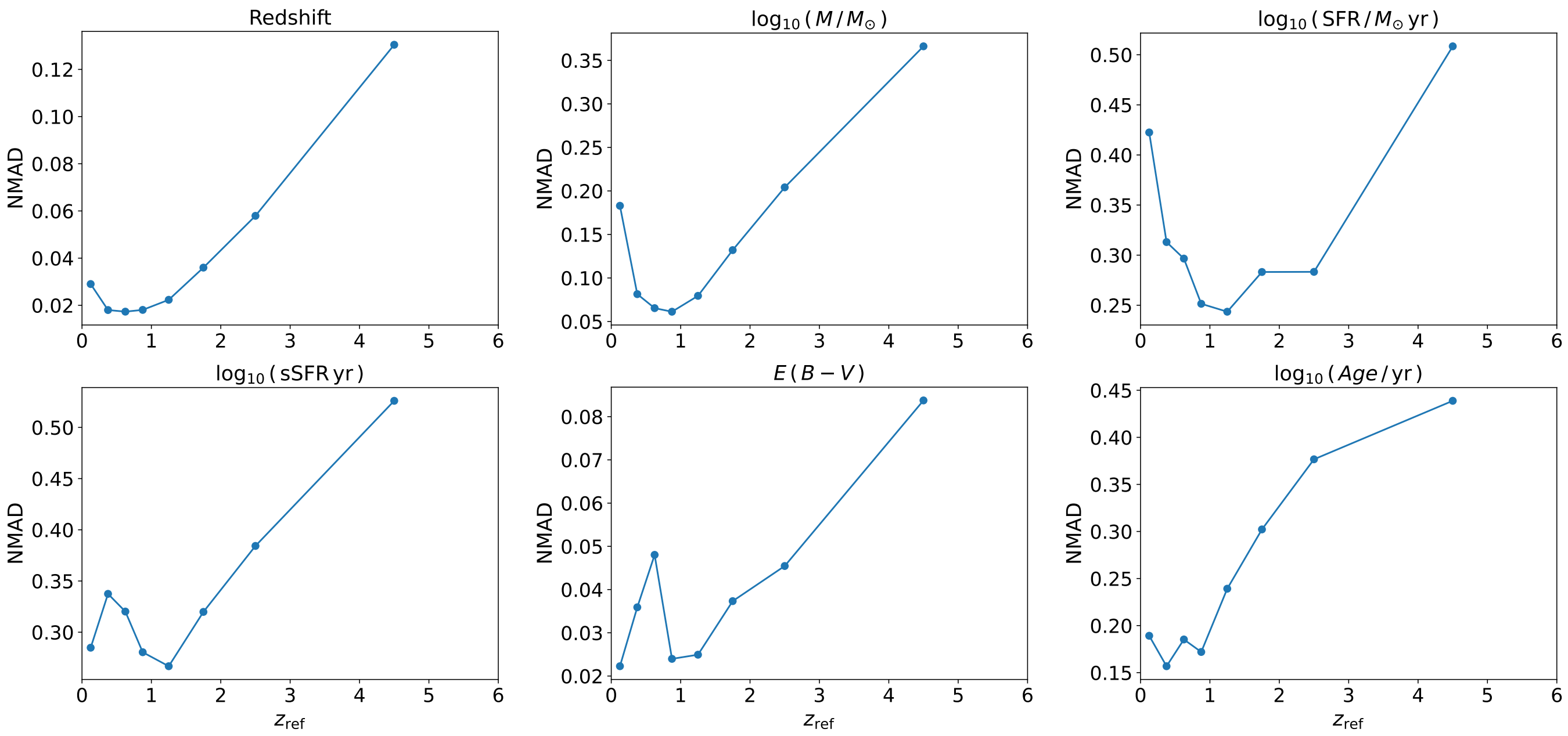

**Fig. A.4.** Example of how the NMAD metric values vary with redshift. For this test, we used the Case 4 data configuration with the Int Wide catalogue. The NMAD metric was calculated after using the ground truth redshift labels to bin the data, with bin edges chosen as follows: 0, 0.25, 0.5, 0.75, 1.0, 1.5, 2.0, 3.0, 6.0.